\documentclass[twocolumn,showpacs,aps,prd]{revtex4}
\usepackage{graphicx}
\usepackage{epsf}  

\newcommand{\BaBarYear}    {06}
\newcommand{\BaBarNumber}  {014}
\newcommand{\SLACPubNumber} {11821}

\input pubboard/babarsym

\def\BRbztodstdstCV {\ensuremath{8.1}}
\def\BRbztodstdstStat {\ensuremath{\pm 0.6 \pm}}
\def\BRbztodstdstSyst {\ensuremath{1.0}}
\def\BRbztodstdstMant{\ensuremath{\BRbztodstdstCV \BRbztodstdstStat \BRbztodstdstSyst}}
\def\BRbztodstdstNum {\ensuremath{(\BRbztodstdstMant)\times 10^{-4}}}

\def\BRbztodstdCV {\ensuremath{5.7}}
\def\BRbztodstdStat {\ensuremath{\pm 0.7 \pm}}
\def\BRbztodstdSyst {\ensuremath{0.7}}
\def\BRbztodstdMant {\ensuremath{\BRbztodstdCV \BRbztodstdStat \BRbztodstdSyst}}
\def\BRbztodstdNum {\ensuremath{(\BRbztodstdMant)\times 10^{-4}}}

\def\BRbztoddCV {\ensuremath{2.8}}
\def\BRbztoddStat {\ensuremath{\pm 0.4 \pm}}
\def\BRbztoddSyst {\ensuremath{0.5}}
\def\BRbztoddMant {\ensuremath{\BRbztoddCV \BRbztoddStat \BRbztoddSyst}}
\def\BRbztoddNum {\ensuremath{(\BRbztoddMant)\times 10^{-4}}}

\def\BRbptodstdstzCV {\ensuremath{8.1}}
\def\BRbptodstdstzStat {\ensuremath{\pm 1.2 \pm}}
\def\BRbptodstdstzSyst {\ensuremath{1.2}}
\def\BRbptodstdzCV {\ensuremath{3.6}}
\def\BRbptodstdzStat {\ensuremath{\pm 0.5 \pm}}
\def\BRbptodstdzSyst {\ensuremath{0.4}}
\def\BRbptoddstzCV {\ensuremath{6.3}}
\def\BRbptoddstzStat {\ensuremath{\pm 1.4 \pm}}
\def\BRbptoddstzSyst {\ensuremath{1.0}}
\def\BRbptoddzCV {\ensuremath{3.8}}
\def\BRbptoddzStat {\ensuremath{\pm 0.6 \pm}}
\def\BRbptoddzSyst {\ensuremath{0.5}}
\def\BRbptodstdstzMant {\ensuremath{\BRbptodstdstzCV \BRbptodstdstzStat \BRbptodstdstzSyst}}
\def\BRbptodstdzMant {\ensuremath{\BRbptodstdzCV \BRbptodstdzStat \BRbptodstdzSyst}}
\def\BRbptoddstzMant {\ensuremath{\BRbptoddstzCV \BRbptoddstzStat \BRbptoddstzSyst}}
\def\BRbptoddzMant {\ensuremath{\BRbptoddzCV \BRbptoddzStat \BRbptoddzSyst}}
\def\BRbptodstdstzNum {\ensuremath{(\BRbptodstdstzMant)\times 10^{-4}}}
\def\BRbptodstdzNum {\ensuremath{(\BRbptodstdzMant)\times 10^{-4}}}
\def\BRbptoddstzNum {\ensuremath{(\BRbptoddstzMant)\times 10^{-4}}}
\def\BRbptoddzNum {\ensuremath{(\BRbptoddzMant)\times 10^{-4}}}

\def\BRbztodstzdstzCV {\ensuremath{-1.3}}
\def\BRbztodstzdstzStat {\ensuremath{\pm 1.1 \pm}}
\def\BRbztodstzdstzSyst {\ensuremath{0.4}}
\def\BRbztodstzdzCV {\ensuremath{1.0}}
\def\BRbztodstzdzStat {\ensuremath{\pm 1.1 \pm}}
\def\BRbztodstzdzSyst {\ensuremath{0.4}}
\def\BRbztodzdzCV {\ensuremath{-0.1}}
\def\BRbztodzdzStat {\ensuremath{\pm 0.5 \pm}}
\def\BRbztodzdzSyst {\ensuremath{0.2}}

\def\LIMbztodstzdstzMant {\ensuremath{0.9}}
\def\LIMbztodstzdzMant {\ensuremath{2.9}}
\def\LIMbztodzdzMant {\ensuremath{0.6}}
\def\LIMbztodstzdstzNum {\ensuremath{\LIMbztodstzdstzMant \times 10^{-4}}}
\def\LIMbztodstzdzNum {\ensuremath{\LIMbztodstzdzMant \times 10^{-4}}}
\def\LIMbztodzdzNum {\ensuremath{\LIMbztodzdzMant \times 10^{-4}}}

\def\ACPbztodstdCV {\ensuremath{0.03}}
\def\ACPbztodstdStat {\ensuremath{\pm 0.10 \pm}}
\def\ACPbztodstdSyst {\ensuremath{0.02}}
\def\ACPbptodstdstzCV {\ensuremath{-0.15}}
\def\ACPbptodstdstzStat {\ensuremath{\pm 0.11 \pm}}
\def\ACPbptodstdstzSyst {\ensuremath{0.02}}
\def\ACPbptodstdzCV {\ensuremath{-0.06}}
\def\ACPbptodstdzStat {\ensuremath{\pm 0.13 \pm}}
\def\ACPbptodstdzSyst {\ensuremath{0.02}}
\def\ACPbptoddstzCV {\ensuremath{0.13}}
\def\ACPbptoddstzStat {\ensuremath{\pm 0.18 \pm}}
\def\ACPbptoddstzSyst {\ensuremath{0.04}}
\def\ACPbptoddzCV {\ensuremath{-0.13}}
\def\ACPbptoddzStat {\ensuremath{\pm 0.14 \pm}}
\def\ACPbptoddzSyst {\ensuremath{0.02}}
\def\ACPbztodstdNum {\ensuremath{\ACPbztodstdCV \ACPbztodstdStat \ACPbztodstdSyst}}
\def\ACPbptodstdstzNum {\ensuremath{\ACPbptodstdstzCV \ACPbptodstdstzStat \ACPbptodstdstzSyst}}
\def\ACPbptodstdzNum {\ensuremath{\ACPbptodstdzCV \ACPbptodstdzStat \ACPbptodstdzSyst}}
\def\ACPbptoddstzNum {\ensuremath{\ACPbptoddstzCV \ACPbptoddstzStat \ACPbptoddstzSyst}}
\def\ACPbptoddzNum {\ensuremath{\ACPbptoddzCV \ACPbptoddzStat \ACPbptoddzSyst}}

\def\Dstarp     {\ensuremath{D^{*+}}\xspace}
\def\Dstarm     {\ensuremath{D^{*-}}\xspace}
\def\Dp         {\ensuremath{D^+}\xspace}
\def\Dm         {\ensuremath{D^-}\xspace}

\def\Bztodstzdstz {\ensuremath{\Bz \to \Dstarz \Dstarzb}\xspace}

\def\Bchtodstzdst {\ensuremath{\Bp \to \Dstarzb \Dstarp}\xspace}

\def\Kpipi {\ensuremath{K \pi \pi}\xspace}

\def\DeltaE     {\ensuremath{\Delta E}\xspace}

\def\masslik    {\ensuremath{{\cal L}_{\rm Mass}}\xspace}

\long\def\inst#1{\par\nobreak\kern 4pt\nobreak
    {\it #1}\par\vskip 10pt plus 3pt minus 3pt}

\begin{document}

\preprint{\babar-PUB-\BaBarYear/\BaBarNumber} 
\preprint{SLAC-PUB-\SLACPubNumber} 

\begin{flushleft}
\babar-PUB-\BaBarYear/\BaBarNumber \\
SLAC-PUB-\SLACPubNumber \\
\end{flushleft}

\title{ 
\Large \bf\boldmath 
Measurement of Branching Fractions and \CP-Violating Charge Asymmetries
for \B\ Meson Decays to $D^{(*)}\Dbar^{(*)}$, and Implications for the CKM Angle $\gamma$ 
}

%
\author{B.~Aubert}
\author{R.~Barate}
\author{M.~Bona}
\author{D.~Boutigny}
\author{F.~Couderc}
\author{Y.~Karyotakis}
\author{J.~P.~Lees}
\author{V.~Poireau}
\author{V.~Tisserand}
\author{A.~Zghiche}
\affiliation{Laboratoire de Physique des Particules, F-74941 Annecy-le-Vieux, France }
% ...
\collaboration{The \babar\ Collaboration}
\noaffiliation

\begin{abstract}
We present measurements of the branching fractions and charge
asymmetries of \B\ decays to all $D^{(*)}\Dbar^{(*)}$ modes.
Using 232 million \BB\ pairs recorded on the $\Upsilon (4S)$ resonance
by the \babar\ detector at the \epem asymmetric $B$ factory \pep2 at the Stanford Linear Accelerator Center,
we measure the branching fractions 
\begin{eqnarray}
\mathcal{B}(\Bz \to D^{*+}D^{*-}) & = & \BRbztodstdstNum, \nonumber\\
\mathcal{B}(\Bz \to D^{*\pm}D^{\mp}) & = & \BRbztodstdNum, \nonumber\\
\mathcal{B}(\Bz \to D^{+}D^{-}) & = & \BRbztoddNum, \nonumber\\
\mathcal{B}(\Bp \to D^{*+}\Dbar^{*0}) & = & \BRbptodstdstzNum, \nonumber\\
\mathcal{B}(\Bp \to D^{*+}\Dbar^{0}) & = & \BRbptodstdzNum, \nonumber\\
\mathcal{B}(\Bp \to D^{+}\Dbar^{*0}) & = & \BRbptoddstzNum, \nonumber\\
\mbox{and}\; \mathcal{B}(\Bp \to D^{+}\Dbar^{0}) & = & \BRbptoddzNum, \nonumber
\end{eqnarray}
where in each case the first uncertainty is statistical and the second systematic.  We also determine the limits
\begin{eqnarray}
\mathcal{B}(\Bz \to D^{*0}\Dbar^{*0}) & < & \LIMbztodstzdstzNum,\nonumber\\ 
\mathcal{B}(\Bz \to D^{*0}\Dbar^{0}) & < & \LIMbztodstzdzNum,\nonumber\\ 
\mbox{and}\; \mathcal{B}(\Bz \to D^{0}\Dbar^{0}) & < & \LIMbztodzdzNum,\nonumber
\end{eqnarray}
each at 90\% confidence level.  All decays above denote either member of a charge
conjugate pair.  We also determine the \CP-violating charge asymmetries 
\begin{eqnarray}
\mathcal{A}(\Bz \to D^{*\pm}D^{\mp}) & = & \ACPbztodstdNum,\nonumber\\
\mathcal{A}(\Bp \to D^{*+}\Dbar^{*0}) & = & \ACPbptodstdstzNum, \nonumber\\
\mathcal{A}(\Bp \to D^{*+}\Dbar^{0}) & = & \ACPbptodstdzNum,\nonumber\\
\mathcal{A}(\Bp \to D^{+}\Dbar^{*0}) & = & \ACPbptoddstzNum,\nonumber\\ 
\mbox{and}\; \mathcal{A}(\Bp \to D^{+}\Dbar^{0}) & = & \ACPbptoddzNum.\nonumber
\end{eqnarray}
Additionally, when we combine these results with information from time-dependent \CP asymmetries in
$\Bz \to D^{(*)+}D^{(*)-}$ decays and
world-averaged branching fractions of $B$ decays to $D_s^{(*)}\Dbar^{(*)}$ modes,
we find the CKM phase $\gamma$ is favored to lie in the range $[0.07-2.77]$ radians (with a +0
or $+\pi$ radians ambiguity) at 68\% confidence level.
\end{abstract}


\maketitle

\newpage

\setcounter{footnote}{0}

\section{Introduction}\label{sec:intro}

We report on measurements of branching fractions of neutral and charged $B$-meson decays to the ten double-charm
final states $D^{(*)}\Dbar^{(*)}$.  For the four charged $B$ decays to $D^{(*)}\Dbar^{(*)}$ and for neutral
$B$ decays to $\Dstarpm\Dmp$, we also measure the direct 
\CP-violating time-integrated charge asymmetry
\begin{equation}
\mathcal{A}_{\CP} \equiv \frac{\Gamma^{-} - \Gamma^{+}}{\Gamma^{-} + \Gamma^{+}},
\label{eq:acp}
\end{equation}
where in the case of the charged $B$ decays, the superscript on $\Gamma$ corresponds to the sign of the \Bpm meson, and
for $\Dstarpm\Dmp$, $\Gamma^{+}$ refers to $\Dstarm\Dp$ and $\Gamma^{-}$ to $\Dstarp\Dm$.

\begin{figure}[h]
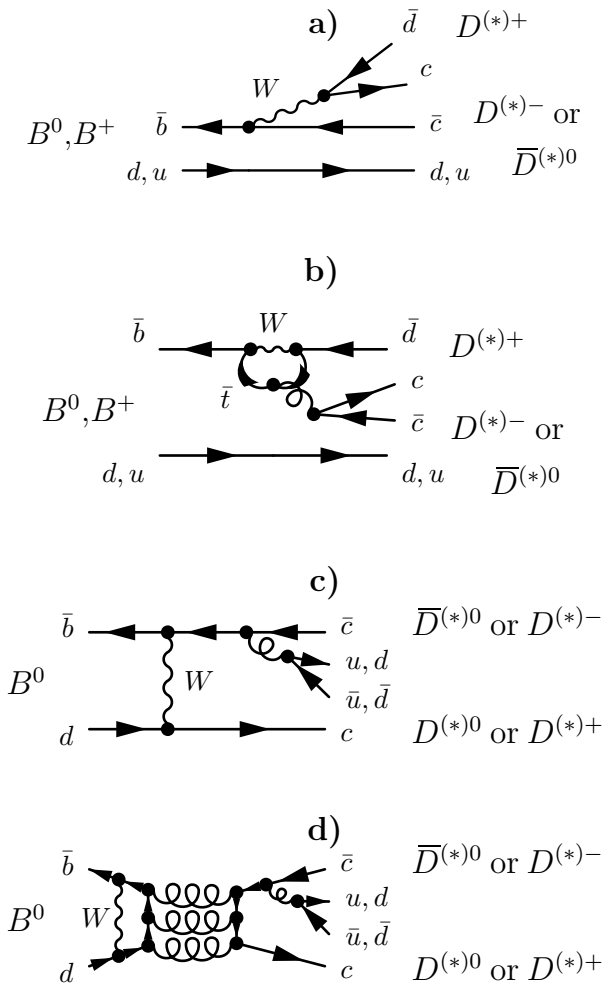
  
\includegraphics{TreeDiagNew2.epsi}

\vspace*{0.9cm}

\scalebox{1.05}{\includegraphics{PenguinDiagNew2.epsi}}

\vspace*{0.9cm}

\scalebox{1.05}{\includegraphics{ExchangeDiag2.epsi}}

\vspace*{0.9cm}

\scalebox{1.05}{\includegraphics{AnnihilationDiag2.epsi}}
\caption{
\label{fig:bddFeynman}
Feynman graphs for $B \to D^{(*)} \Dbar^{(*)}$ decays:
the tree (a) and penguin (b) diagrams are the leading terms for both \Bz $\to D^{(*)+}D^{(*)-}$ and \Bp $\to D^{(*)+}\Dbar^{(*)0}$ 
decays, whereas the exchange (c) and annihilation (d) diagrams (the latter of which is OZI-suppressed) are the lowest-order terms 
for \Bz $\to D^{(*)0}\Dbar^{(*)0}$ decays.}
\end{figure}

In the neutral $B \to D^{(*)+}D^{(*)-}$ decays, the interference of the dominant 
tree diagram (see Fig.~\ref{fig:bddFeynman}a) with the neutral $B$ mixing diagram is sensitive to the Cabibbo-Kobayashi-Maskawa (CKM) phase 
$\beta \equiv \arg \left[\, -V_{\rm cd}^{}V_{\rm cb}^* / V_{\rm td}^{}V_{\rm tb}^*\, \right]$, where $V$ is the CKM quark mixing matrix~\cite{CKM}.
However, the theoretically uncertain contributions of penguin diagrams (Fig.~\ref{fig:bddFeynman}b) with different weak phases are potentially
significant and may shift both the observed \CP asymmetries and the branching fractions by amounts that depend on the ratios of the penguin to 
tree contributions and their relative phases.  A number of theoretical estimates exist for the resulting values of the branching fractions and \CP
asymmetries~\cite{gronauetc,rosner,sandaxing,phamxing,xing2}.

The penguin-tree interference in neutral and charged $B \to D^{(*)} \Dbar^{(*)}$ decays can provide sensitivity to the angle
$\gamma = \arg \left[\, -V_{\rm ud}^{}V_{\rm ub}^* / V_{\rm cd}^{}V_{\rm cb}^*\, \right]$~\cite{DL,ADL}.  With additional information on the branching fractions of
$B \to D^{(*)}_s \Dbar^{(*)}$ decays,
the weak phase may be extracted, assuming SU(3) flavor symmetry between $B \to D^{(*)} \Dbar^{(*)}$ and $B \to D^{(*)}_s \Dbar^{(*)}$.
For this analysis, we assume that the breaking of SU(3) can be parametrized via the ratios of decay
constants $f_{D_s^{(*)}}/f_{D^{(*)}}$, which are quantities that can be
determined either with lattice QCD or from experimental measurements~\cite{lattice}.

In addition to presenting measurements of the \Bz $\to D^{(*)+}D^{(*)-}$ and \Bp $\to D^{(*)+} \Dbar^{(*)0}$ branching fractions, and
the \CP-violating charge asymmetries for the latter modes and for \Bz $\to D^{*\pm}D^{\mp}$,
we search for the color-suppressed decay modes \Bz $\to D^{(*)0}\Dbar^{(*)0}$,
which have not been previously measured, and determine limits on those branching fractions~\cite{CC}.  If observed, the decays \Bz $\to D^{(*)0}\Dbar^{(*)0}$
would provide evidence of $W$-exchange or annihilation contributions (see Fig.~\ref{fig:bddFeynman}c,\ref{fig:bddFeynman}d).
In principle, these decays could also provide sensitivity to the CKM phase $\beta$ if sufficient data were 
available.
By combining all of these results with information from time-dependent \CP asymmetries in
$\Bz \to D^{(*)+}D^{(*)-}$ decays and
world-averaged branching fractions of $B$ decays to $D_s^{(*)}\Dbar^{(*)}$ modes, we determine the implications for $\gamma$ using the method of Refs.~\cite{DL,ADL}.

\section{Detector and Data} \label{sec:detector}

The results presented in this paper are based on data collected
with the \babar\ detector~\cite{BABARNIM}
at the PEP-II asymmetric-energy $e^+e^-$ collider~\cite{pep}
located at the Stanford Linear Accelerator Center. 
The integrated luminosity
is 210.5~\invfb, corresponding to 231.7 million \BB\ pairs, 
recorded at the $\Upsilon (4S)$ resonance
(``on-peak'', at a center-of-mass (c.m.) energy $\sqrt{s}=10.58\ \gev$).

The asymmetric beam configuration in the laboratory frame
provides a boost of $\beta\gamma = 0.56$ to the $\Upsilon(4S)$.  
Charged particles are detected and their momenta measured by the
combination of a silicon vertex tracker (SVT), consisting of five layers of double-sided detectors,
and a 40-layer central drift chamber (DCH),
both operating in the 1.5-T magnetic field of a solenoid.  
For tracks with transverse momentum greater than 120~\mevc, the DCH provides the
primary charged track finding capability.
The SVT provides complementary standalone track finding for tracks of lower momentum, allowing for
reconstruction of charged tracks with transverse momentum $p_T$
as low as 60 \mevc, with efficiencies in excess of 85\%.  
This ability to reconstruct tracks with low $p_T$ efficiently is  
necessary for reconstruction of the slow charged pions from
\Dstarp $\to \Dz\pip$ decays in $B \to D^{(*)} \Dbar^{(*)}$ signal events. 
The transverse momentum resolution for the combined tracking system is 
$\sigma_{p_T}/p_T=0.0013p_T + 0.0045$, where 
$p_T$ is measured in \gevc.  
Photons are detected and their energies measured by a CsI(Tl) electromagnetic
calorimeter (EMC).  The photon energy resolution is 
$\sigma_{E}/E = \left\{2.3 / E(\gev)^{1/4} \oplus 1.4 \right\}\%$, 
and their angular resolution with respect to the interaction point is
$\sigma_{\theta} = \mbox{(4.2 mrad)}/\sqrt{E(\gev)}$. 
The measured $\pi^0$ mass resolution for $\piz$'s with
laboratory momentum in excess of 1 \gevc\ is approximately 6 \mevcc.

Charged-particle identification (PID) is provided by 
an internally reflecting ring-imaging
Cherenkov light detector (DIRC) covering the central region,
and the most probable energy loss (\dedx) in the tracking devices.
The Cherenkov angle resolution of the DIRC is measured to be 2.4~mrad, which
provides over $5\sigma$ separation between charged kaons and pions at
momenta of less than $2~\gevc$.  The $\dedx$ resolution 
from the drift chamber is typically about $7.5\%$ for pions.  
Additional information
to identify and reject electrons and muons is provided
by the EMC and detectors embedded between the steel plates of
the magnetic flux return (IFR).

\section{Candidate Reconstruction and {\boldmath $B$} Meson Selection}
\label{sec:eventsel}

Given the high multiplicity of the final states studied, very high
combinatorial background levels are expected. Selection criteria (described
in Sec.~III A--E)  are designed to minimize the expected statistical error on the 
$B$ branching fractions (as described in Sec.~III F).
A {\tt GEANT4}-based~\cite{GEANT4} Monte Carlo (MC) simulation
of the material composition and the instrumentation response of the \babar\ detector is
used to optimize signal selection criteria and evaluate signal detection efficiency.
We retain sufficient sidebands in the discriminating
variables to characterize the background in subsequent fits.

\subsection{Charged track and {\boldmath \KS} selection}
\label{sec:track_sele}

Charged particle tracks are selected via pattern recognition algorithms
using measurements from the SVT and DCH detectors.
We additionally require all charged-particle tracks (except for those from 
$\KS\ra\pip\pim$ decays) to
originate within 10 cm along the beam axis and 1.5 cm in the plane perpendicular to the beam axis of the center of the
beam crossing region.  To ensure a well-measured momentum, all charged-particle tracks except
those from $\KS\ra\pip\pim$ decays and \pip from $\Dstarp \to \Dz\pip$ decays must also
be reconstructed from at least 12 measurements in the DCH.
All tracks that meet these criteria are considered as charged pion candidates.

Tracks may be identified as kaons based on a likelihood selection developed from 
Cherenkov angle and \dedx information from the DIRC and tracking detectors respectively.
For the typical laboratory momentum spectrum of the signal kaons, this  
selection has an efficiency of about 85\% and a purity of 
greater than 98\%, as determined from control samples of $\Dstarp \to \Dz\pip$,
$D^0\to K^-\pip$ decays.  

We require $\KS \to \pip\pim$ candidates to have an invariant mass within 15 \mevcc of the nominal \KS
mass~\cite{PDG2004}.  The probability that the two daughter tracks originate from the same 
point in space must be greater than 0.1\%.  The
transverse flight distance of the \KS from the primary event vertex 
must be both greater than $3\sigma$ from zero (where $\sigma$ is the measured uncertainty on the transverse flight length) and
also greater than 2 mm.

\subsection{Photon and {\boldmath \piz} selection}
\label{sec:gampiz_sele}

Photons are reconstructed from energy deposits in the electromagnetic
calorimeter which are not associated with a charged track.  To reject backgrounds
from electronics noise, machine background, and hadronic interactions in the EMC, we require that all photon
candidates have an energy greater than 30 \mev in the laboratory frame and to have a 
lateral shower shape consistent with that of a photon.
Neutral pions are reconstructed from pairs of photon candidates whose energies in the laboratory frame sum to more than 200 \mev.
The \piz candidates must have an invariant mass between 115 and 150~\mevcc.  The 
\piz candidates that meet these criteria, when combined with other tracks or neutrals to form $B$ candidates, 
are then constrained to originate from their expected decay points,
and their masses are constrained to the nominal value~\cite{PDG2004}.
This procedure improves the mass and energy resolution of the parent
particles.

\begin{figure*}[t]
\begin{center}
\scalebox{0.85}{\includegraphics{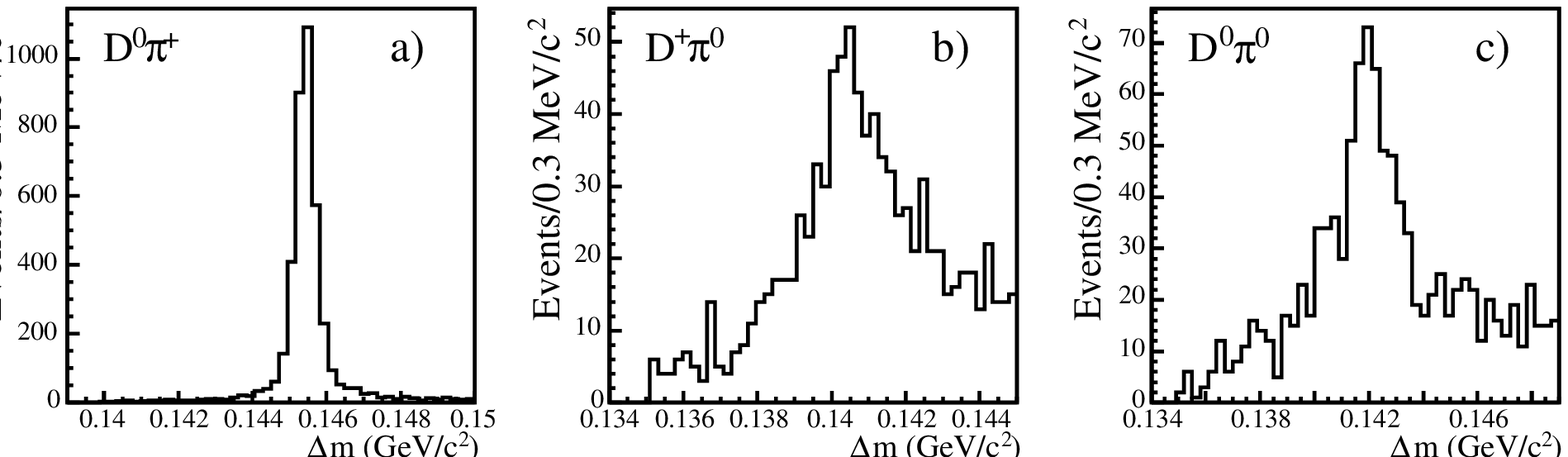}}
\end{center}
\vspace*{-0.6cm}
\caption{
\label{fig:deltam}
Distributions of $\Delta m$ in the full data sample for three \Dstar decay modes.
Plot a) shows $\Delta m(\Dstarp - \Dz)$ for $\Dstarp \to \Dz\pip$ decays where
\Dz decays to $\Km\pip$.  Plot b) shows $\Delta m(\Dstarp - \Dp)$ for
$\Dstarp \to \Dp\piz$ decays where \Dp decays to $\Km\pip\pip$.  Plot c)
shows $\Delta m(\Dstarz - \Dz)$ for $\Dstarz \to \Dz\piz$ decays where \Dz decays
to $\Km\pip$.  
Nominal values for $\Delta m$ are 145.4 \mevcc, 140.6 \mevcc, and 
142.1 \mevcc for the three cases respectively~\cite{PDG2004}.
}
\end{figure*}

\subsection{Event selection}
\label{sec:event_sele}

We select \BB events by applying criteria on the
track multiplicity and event topology. 
At least three reconstructed
tracks, each with transverse momentum greater than 100 \mevc, are required in the
laboratory polar angle region $0.41 < \theta_{\rm lab} < 2.54$.  
The event must have a total measured energy in the laboratory frame greater than 4.5~\gev to reject beam-related background.  
The ratio of Fox-Wolfram moments $H_2/H_0$~\cite{FoxW} is a parameter between 0 (for ``perfectly spherical'' events) and 1 (for ``perfectly jet-like'' events), 
and we require this ratio to be less than 0.6 for each event, in order to help reject non-\BB background.
This criterion rejects between 30 and 50 percent of non-\BB background (depending on the decay mode), while keeping almost all of the signal decays.

\subsection{{\boldmath $D$} and {\boldmath \Dstar} meson selection}
\label{sec:dmeson_sele}

We reconstruct \Dz mesons in the four decay modes $\Dz \to \Km\pi^{+}$, $\Dz \to \Km\pip\piz$, $\Dz \to \Km\pip\pim\pip$,
and $\Dz \to \KS\pip\pim$, and \Dp mesons in the two decay modes $\Dp \to \Km\pip\pip$ and $\Dp \to \KS\pip$.
We require \Dz and \Dp candidates to have 
reconstructed masses 
within $\pm20$ \mevcc of their nominal masses~\cite{PDG2004}, except for 
$\Dz \to \Km\pip\piz$, for which we require $\pm40$ \mevcc due to the poorer resolution for modes containing
\piz's.  
These criteria correspond to approximately $2.5 \sigma$ of the respective mass resolutions.
The $\Dz \to \Km\pip\piz$ decays must also satisfy a criterion on the reconstructed invariant
masses of the $\Km\pip$ and $\Km\piz$ pairs: the combination of reconstructed invariant masses must lie at a point 
in the $\Km\pip\piz$ Dalitz plot~\cite{Dalitz} for  
which the expected density normalized to the maximum density 
(``Dalitz weight'') is at least 6\%.
Additionally, the daughters of \Dz and \Dp candidates 
must have a probability of originating from a common point in space greater than 0.1\%, and are then
constrained both to originate from that common spatial point and to have their respective nominal invariant masses.

\begin{figure*}[t]
\begin{center}
\includegraphics[width=0.8\textwidth]{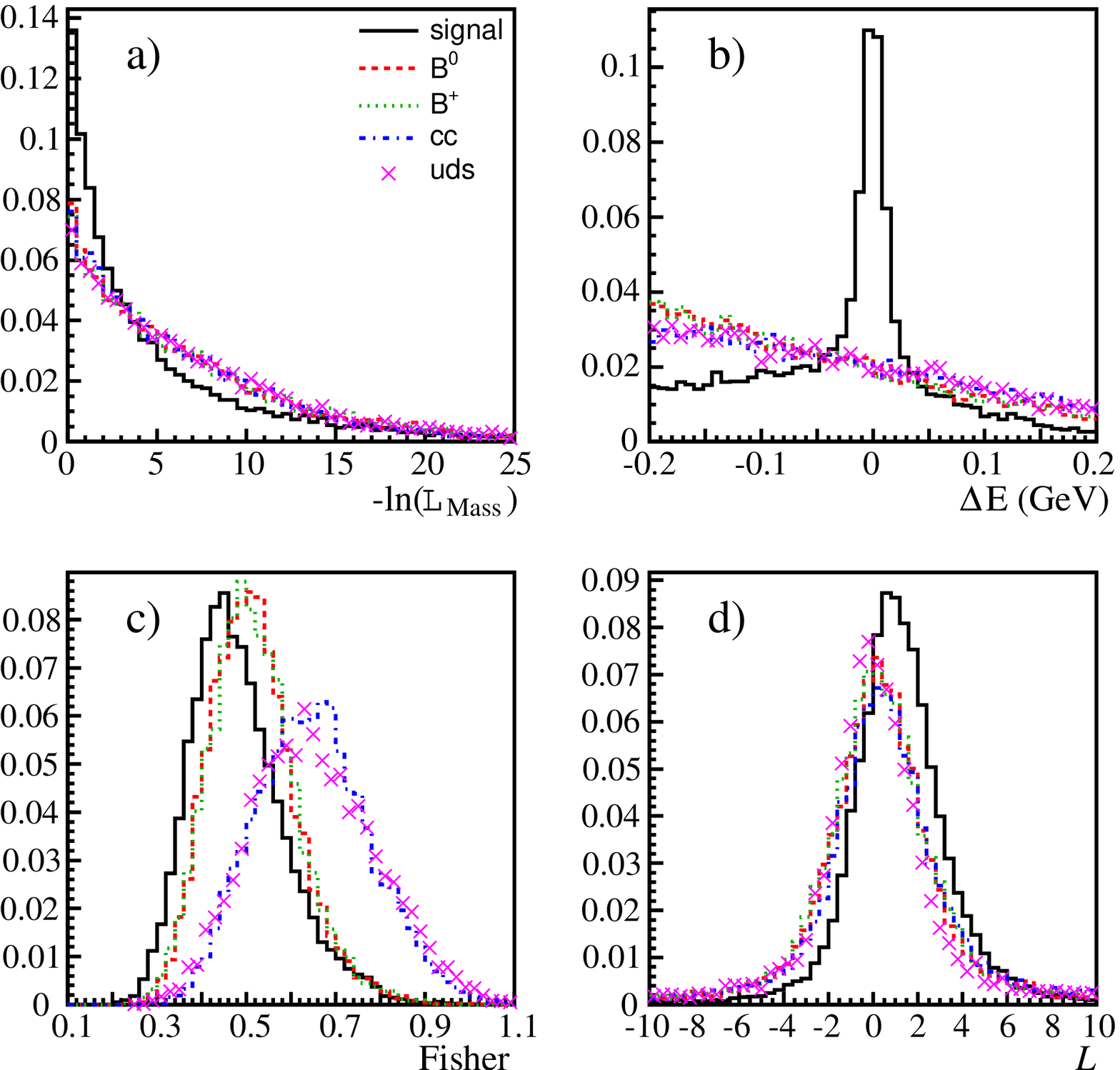}
\end{center}
\vspace*{-0.6cm}  
\caption{
\label{fig:selvars}
Distributions of signal selection variables: a) the likelihood variable $-\ln(\masslik)$,
b) the $\Delta E$ variable,
c) the Fisher discriminant $\mathcal{F}$, and
d) the $D$-meson flight length variable $L$,
each for the representative signal mode \Bz $\to \Dz\Dzb$, and
for the corresponding combinatorial background from \BzBzb, \BpBm, \ccbar, and (\uubar+\ddbar+\ssbar) 
MC simulated decays respectively.  In each plot, the component distributions are normalized to have the same area below the curves.
}
\end{figure*}

Candidate \Dstarp and \Dstarz mesons are reconstructed in the decay modes $\Dstarp \to \Dz\pip$, $\Dstarp \to \Dp\piz$,
$\Dstarz \to \Dz\piz$, and $\Dstarz \to \Dz\gamma$, using pairs of selected \Dz, \Dp, \piz, \pip, and $\gamma$ candidates.
The \pip from $\Dstarp \to \Dz\pip$ decays is additionally required to have a c.m.~momentum of less than 450 \mevc.  
Candidate \piz mesons from $\Dstarp \to \Dp\piz$ and $\Dstarz \to \Dz\piz$ are required to have c.m.~momenta 
$p^{*}$ in the range $70 < p^{*} < 450$~\mevc.  Photons from $\Dstarz \to \Dz\gamma$ decays are required
to have energies in the laboratory frame greater than 100 \mev and c.m.~energies less than 450 \mev.  The \Dstar
daughter particles are constrained to originate from a common point in space.  After this constraint is applied, the mass differences
$\Delta m$ of the reconstructed masses of the \Dstar and $D$ candidates are required to be within 
the ranges shown in Table~\ref{tab:deltaMcuts}.
\begin{table}[h]
\caption{\label{tab:deltaMcuts}
Allowed $\Delta m$(\Dstar - $D$) ranges for the four \Dstar decay modes.
}
\begin{center}
\begin{tabular}{lcc}
\hline \hline
                        & Minimum             & Maximum             \\
Mode                    & $\Delta m$ (\mevcc) & $\Delta m$ (\mevcc) \\
\hline
$\Dstarp \to \Dz\pip$   &  $139.6$                    &  $151.3$                    \\
$\Dstarp \to \Dp\piz$   &  $135.0$                    &  $146.3$                    \\
$\Dstarz \to \Dz\piz$   &  $135.0$                    &  $149.3$                    \\
$\Dstarz \to \Dz\gamma$ &  $100.0$                    &  $170.0$                    \\
\hline \hline
\end{tabular}
\end{center}
\end{table}
As shown in Fig.~\ref{fig:deltam}, the excellent resolution in $\Delta m$ for signal candidates makes the
$\Delta m$ requirement a very powerful 
criterion to reject background (see next section), especially for decay modes containing a $\Dstarp \to \Dz\pip$.

\subsection{Variables used for {\boldmath $B$} meson selection}
\label{sec:bmesonvar_sele}

A $B$-meson candidate is constructed by combining two $D^{(*)}$ candidates that have both passed the selection
criteria described previously.  The pairs of $D^{(*)}$ candidates are constrained to originate from the
same point in space.  We form a likelihood variable, $\masslik$, that is defined by a product
of Gaussian distributions for each $D$ mass and $\Dstar-D$ mass difference.
\begin{table*}[t]
\caption{\label{tab:globalcuts}Expected values of the branching fractions $\mathcal{B}$ for each $B \to D^{(*)} \Dbar^{(*)}$ decay mode, which are used for the purpose of determining selection
                               criteria that minimize the expected uncertainty on the measured branching fraction for each mode; also, optimized 
                               $\mathcal{F}$ and $L$ selection criteria for each mode.  
                               An ``---'' indicates no cut is made in $\mathcal{F}$ or $L$ for that 
                               decay mode.}
\begin{center}
\scalebox{1.0}{
\begin{tabular}{lccc}
\hline \hline
Mode                         &  Expected $\mathcal{B}$ & $\mathcal{F}_{\rm min}$ &  $L_{\rm min}$   \\
\hline 
\Bz $\to D^{*+}D^{*-}$       &  $8.3 \times 10^{-4}$   &    ---        &  ---   \\
\Bz $\to D^{*\pm}D^{\mp}$    &  $8.8 \times 10^{-4}$   &    ---        &  ---   \\
\Bz $\to D^{+}D^{-}$         &  $3.0 \times 10^{-4}$   &      0.62     &  $1.3$  \\
\Bz $\to D^{*0}\Dstarzb$ &  $1.0 \times 10^{-5}$   &      0.60     &  $-1.6$  \\
\Bz $\to D^{*0}\Dzb$  &  $1.0 \times 10^{-5}$   &      0.53     &  $-0.4$  \\
\hline
\end{tabular}
}
\hspace{1cm}
\scalebox{1.0}{
\begin{tabular}{llccc}
\hline \hline
Mode                         &  Expected $\mathcal{B}$ & $\mathcal{F}_{\rm min}$ &  $L_{\rm min}$   \\
\hline 
\Bz $\to D^{0}\Dzb$   &  $1.0 \times 10^{-5}$   &      0.47     &  $-0.4$  \\
\Bp $\to D^{*+}\Dstarzb$ &  $1.0 \times 10^{-3}$  &      0.60     &  ---   \\
\Bp $\to D^{*+}\Dzb$  &  $4.4 \times 10^{-4}$   &      0.53     &  $-1.3$  \\
\Bp $\to D^{+}\Dstarzb$  &  $4.4 \times 10^{-4}$   &      0.53     &  $ 0.0$  \\
\Bp $\to D^{+}\Dzb$   &  $3.0 \times 10^{-4}$   &      0.53     &  $ 0.5$  \\
\hline
\end{tabular}
}
\end{center}
\end{table*}

For example, in the decay $\Bz \to \Dstarp\Dstarm$, $\masslik$ is the product of four terms:
Gaussian distributions for each $D$ mass and double Gaussian (\textit{i.e.}~the sum of two Gaussian distributions) terms
for each $\Delta m$ term (the $\Dstar - D$ mass difference).  Defining $G(x;\mu,\sigma)$ as a normalized
Gaussian distribution where $x$ is the independent variable, $\mu$ is the mean, and $\sigma$ is the resolution,
$\masslik$ for $\Bz \to \Dstarp\Dstarm$ decays is defined as:
\begin{widetext}
\begin{eqnarray}
\label{eq:chi2likedstd}
\masslik & = & G(m_D;m_{D_{\rm PDG}},\sigma_{m_D}) \times
                      G(m_{\Dbar};m_{\Dbar_{\rm PDG}},\sigma_{m_{\Dbar}}) \times
                  \nonumber \\
&  & \qquad \left[ f_{\rm core} G(\Delta m_{\Dstarp};\Delta m_{\Dstarp_{\rm PDG}},\sigma_{\Delta m_{\rm core}})
                     + (1 - f_{\rm core}) G(\Delta m_{\Dstarp};\Delta m_{\Dstarp_{\rm PDG}},\sigma_{\Delta m_{\rm tail}})  \right]
           \times  \nonumber \\
&  & \qquad \left[ f_{\rm core} G(\Delta m_{\Dstarm};\Delta m_{\Dstarm_{\rm PDG}},\sigma_{\Delta m_{\rm core}})
                     + (1 - f_{\rm core}) G(\Delta m_{\Dstarm};\Delta m_{\Dstarm_{\rm PDG}},\sigma_{\Delta m_{\rm tail}})  \right],
\end{eqnarray}
\end{widetext}
where the subscript ``PDG'' refers to the nominal value~\cite{PDG2004}, and all reconstructed masses and uncertainties are determined before mass constraints are applied.  
For $\sigma_{m_D}$, we use errors calculated 
candidate-by-candidate.  The parameter $f_{\rm core}$ is the ratio of the area of the core Gaussian to the total area of the double
Gaussian distribution.  This, along with $\sigma_{\Delta m_{\rm core}}$ and $\sigma_{\Delta m_{\rm tail}}$, is determined separately 
for each of the four \Dstar decay modes given above, using MC simulation of signal events that is calibrated to inclusive samples
of the \Dstar decay modes in data.  
For each of the $B$ decay modes, a higher value of \masslik tends to indicate a greater signal likelihood.  
The distributions of $-\ln(\masslik)$ for the representative signal mode \Bz $\to \Dz\Dzb$ and
for the corresponding combinatorial background from generic \BzBzb, \BpBm, \ccbar, and (\uubar+\ddbar+\ssbar) decays, are shown in 
Fig.~\ref{fig:selvars}a.
We use  
\masslik in selecting signal candidates, as will be described in the upcoming section.

We also use the two variables for fully-reconstructed $B$ meson selection at the \Y4S energy: the beam-energy-substituted mass 
\mes $\equiv [(s/2 + \vec{p}_i \cdot \vec{p}_B)^2/E_i^2 - \vec{p}_B^2]^{1/2}$, where the initial total \epem four-momentum $(E_i,\vec{p}_i)$
and the $B$ momentum $\vec{p}_B$ are defined in the laboratory frame; and 
$\Delta E \equiv E_B^{\rm cm} - \sqrt{s}/2$ is
the difference between the reconstructed $B$ energy in the c.m.~frame
and its known value.
The normalized distribution of $\Delta E$ for the representative signal mode \Bz $\to \Dz\Dzb$, and
for the corresponding combinatorial background components, 
is shown in Fig.~\ref{fig:selvars}b.

In addition to \masslik, \mes, and $\Delta E$, a Fisher discriminant $\mathcal{F}$~\cite{cleofisher} and a $D$-meson flight length variable $L$ are used
to help separate signal from background.  The Fisher discriminant assists in the suppression of background from continuum events by incorporating 
information from the topology of the event.  The discriminant is formed from the momentum flow into nine polar angular intervals of $10^\circ$ centered on the
thrust axis of the $B$ candidate, the angle of the event thrust axis 
with respect to the beam 
axis ($\theta_{T}$),
and the angle of the $B$ candidate momentum with respect to the beam axis ($\theta_B$):
\begin{equation}
\mathcal{F} \equiv \sum_{i=1}^{11} \alpha_i x_i.
\end{equation}
The values $x_i~(i = 1, ..., 9)$ are the scalar sums of the momenta of all charged
tracks and neutral showers in the polar angle interval $i$, $x_{10}$ is
$|\text{cos}\theta_{T}|$, and $x_{11}$ is $|\text{cos}\theta_B|$.
The coefficients $\alpha_i$ are determined from MC simulation to maximize
the separation between signal and background~\cite{cleofisher}.
The normalized distribution of $\mathcal{F}$ for the representative signal mode \Bz $\to \Dz\Dzb$, and
for the corresponding background components,
is shown in Fig.~\ref{fig:selvars}c.

The flight length variable $L$ that we consider is defined as 
$(\ell_1 + \ell_2)/\sqrt{\sigma_1^2 + \sigma_2^2}$, with the decay lengths $\ell_i$ of the two $D$ mesons defined as
\begin{equation}
{\vec x}_{D_i} = {\vec x}_B + (\ell_i \times {\vec p}_{D_i}) 
\end{equation}
where $\vec{x}_D$ and $\vec{x}_B$ are the measured decay vertices of the $D$ and $B$, respectively, and $\vec{p}_D$ is
the momentum of a $D$.
The $\sigma_i$ are the measured uncertainties on $\ell_i$.
This observable exploits the ability
to distinguish the long $D$ lifetime.  Thus, background events
have an $L$ distribution centered around zero, while
events with real $D$ mesons have a distribution favoring
positive values.
The normalized distribution of $L$ for the representative signal mode \Bz $\to \Dz\Dzb$, and
for the corresponding background components,
is shown in Fig.~\ref{fig:selvars}d.

\subsection{Analysis optimization and signal selection}
\label{sec:analopti_sele}

We combine information from the \masslik, $\Delta E$, $\mathcal{F}$, and $L$ variables to select signal candidates in each decay mode.
The fractional statistical uncertainty on a measured branching fraction is proportional to $\sqrt{(N^s + N^b)}/N^s$, where $N^s$ is the 
number of reconstructed signal events and $N^b$ is the number of background events within the selected signal region for a mode. 
The values $N^s$ and $N^b$ are calculated,
using detailed MC simulation of the signal decay modes as well as of \BB and continuum background decays,
by observing the number of simulated $B$ decay candidates
that satisfy the selection criteria for $-\ln(\masslik)$, $|\Delta E|$, $\mathcal{F}$, and $L$.
We choose criteria which minimize the expected $\sqrt{(N^s + N^b)}/N^s$ for each mode.  Note that to calculate the expected number of signal events $N^s$,
one must assume an expected branching fraction, as well as the ratios of
\BB and continuum events using their relative cross-sections.  These are given, along with the requirements on $\mathcal{F}$ and $L$, in Table~\ref{tab:globalcuts}.

For each possible combination of \Dstarp, \Dstarz, \Dp, and \Dz decay modes, we determine the combination of selection criteria on
$-\ln(\masslik)$ and $|\Delta E|$ that minimizes the overall expected $\sqrt{(N^s + N^b)}/N^s$ for each $B$ decay mode (see Tables~\ref{tab:modekey}, ~\ref{tab:masslikcuts}, 
and~\ref{tab:deltaecuts}).  
The selection criteria for $\mathcal{F}$ and $L$ are chosen, however, 
only for each $B$ decay mode and not separately for each $D^{(*)}$ mode combination.  The restrictiveness of the kaon identification selection
is also optimized separately for each charged and neutral $D^{(*)}$ mode.  

Between 1\% and 34\% of selected $B \to D^{(*)} \Dbar^{(*)}$ events have more than one reconstructed $B$ candidate
that passes all selection criteria in
\masslik, $\Delta E$, $\mathcal{F}$, and $L$, with the largest percentages
occurring in the decay modes \Bz $\to \Dstarz\Dstarzb$ and \Bz $\to \Dz\Dzb$, and the smallest occurring in \Bz $\to \Dstarpm\Dmp$ and \Bz $\to \Dp\Dm$.
In such events, we choose the reconstructed $B$ with the largest value of \masslik as the
signal $B$ candidate.

\begin{table*}[p]
\caption{\label{tab:modekey} Key to mode numbers used in Tables~\ref{tab:masslikcuts} and~\ref{tab:deltaecuts} below.}
\begin{center}
\scalebox{0.90}{
\begin{tabular}{lc}
\hline \hline
Mode                                    &  \#  \\
\hline
\Dstarp $\to (K^-\pi^+)\pi^+$           &  1 \\
\Dstarp $\to (K^-\pi^+\piz)\pi^+$       &  2 \\
\Dstarp $\to (K^-\pi^+\pi^-\pi^+)\pi^+$ &  3 \\
\Dstarp $\to (\KS\pi^+\pi^-)\pi^+$      &  4 \\
\Dstarp $\to (K^-\pi^+\pi^+)\piz$       &  5 \\
\Dstarz $\to (K^-\pi^+)\piz$            &  6 \\
\Dstarz $\to (K^-\pi^+\piz)\piz$        &  7 \\
\Dstarz $\to (K^-\pi^+\pi^-\pi^+)\piz$  &  8 \\
\hline \hline
\end{tabular}
}
\hspace{1cm}
\scalebox{0.90}{
\begin{tabular}{lc}
\hline \hline
Mode                                     & \# \\
\hline
\Dstarz $\to (\KS\pi^+\pi^-)\piz$        &  9 \\
\Dstarz $\to (K^-\pi^+)\gamma$           & 10 \\
\Dp $\to K^-\pip\pip$                    & 11 \\
\Dz $\to \Km\pip$                        & 12 \\
\Dz $\to \Km\pip\piz$                    & 13 \\
\Dz $\to \Km\pip\pim\pip$                & 14 \\
\Dz $\to \KS\pi^+\pi^-$                  & 15 \\
                                         &    \\
\hline \hline
\end{tabular}
}
\vspace*{-0.3cm}
\end{center}
\end{table*}
\begin{table*}[p]
\caption{\label{tab:masslikcuts}Optimized $-\ln(\masslik)$ selection criteria used for all $B \to D^{(*)} \Dbar^{(*)}$ modes.  Selected events in a given mode must have 
                                $-\ln(\masslik)$ less than the given value.  The $D^{(*)}$ decay modes $1-15$ are defined in Table~\ref{tab:modekey} above.
                                Elements with ``---'' above and on the diagonal are modes that are 
                                unused since, due to high backgrounds, they do not help to increase signal sensitivity.}
\begin{center}
\scalebox{0.83}{
\begin{tabular}{ll|ccccc|ccccc|c|cccc}
\hline \hline
          &                    &  \multicolumn{5}{|c|}{\rule[0mm]{0mm}{5mm}\raisebox{0.5mm}[0mm]{\Large $D^{*-}$}} & 
                                  \multicolumn{5}{|c|}{\rule[0mm]{0mm}{5mm}\raisebox{0.5mm}[0mm]{\Large $\Dstarzb$}} &
                                  \multicolumn{1}{|c|}{\rule[0mm]{0mm}{5mm}\raisebox{0.5mm}[0mm]{\Large $D^{-}$}}  &
                                  \multicolumn{4}{|c}{\rule[0mm]{0mm}{5mm}\raisebox{0.5mm}[0mm]{\Large $\Dzb$}} \\
          &                    &   1   &   2   &   3   &   4   &   5   &   6   &   7   &   8   &   9   &  10   &  11   &    12  &   13  &   14   &   15  \\
\hline
          &          1         & 13.0  & 12.0  & 17.3  & 19.8  & 10.5  & 14.6  & 17.5  & 9.2   &  ---  & 8.9   &  8.2  &   8.6  &  8.5  &  8.2   &  8.0  \\
          &          2         &       & 10.6  & 11.0  & 18.3  &  9.5  & 11.5  &  9.8  & 10.7  &  ---  & 8.7   &  8.4  &   7.8  &  ---  &  8.8   & ---   \\
{\Large $D^{*+}$}&   3         &       &       & 11.7  & 11.0  &  9.8  & 11.7  &  9.6  & 10.4  &  ---  & 9.0   &  8.8  &   9.3  &  9.4  &  9.0   & ---   \\
          &          4         &       &       &       &  ---  &  ---  &  ---  &  ---  &  ---  &  ---  & ---   &  9.6  &  15.1  &  9.2  &  ---   & ---   \\
          &          5         &       &       &       &       &  ---  &  8.2  &  ---  &  ---  &  ---  & ---   &  ---  &   6.6  &  ---  &  ---   & ---   \\
\hline
          &          6         &       &       &       &       &       & 12.2  &  8.4  & 9.6   & 7.6   & ---   &  9.9  &   7.6  &  6.7  &  7.2   & ---   \\
          &          7         &       &       &       &       &       &       &  ---  &  ---  &  ---  & ---   &  7.5  &   ---  &  ---  &  ---   & ---   \\
{\Large $D^{*0}$}&   8         &       &       &       &       &       &       &       &  ---  &  ---  & ---   &  9.2  &   ---  &  ---  &  ---   & ---   \\
          &          9         &       &       &       &       &       &       &       &       &  ---  & ---   &  ---  &   5.8  &  ---  &  ---   & ---   \\
                 &  10         &       &       &       &       &       &       &       &       &       & ---   &  ---  &   ---  &  ---  &  ---   & ---   \\
\hline
{\Large $D^{+}$} &  11         &       &       &       &       &       &       &       &       &       &       &  6.0  &   7.3  &  5.8  &  6.5   &  6.2  \\
\hline
                 &  12         &       &       &       &       &       &       &       &       &       &       &       &   ---  &  5.2  &  6.8   & ---   \\
                 &  13         &       &       &       &       &       &       &       &       &       &       &       &        &  ---  &  6.2   & ---   \\
\raisebox{1.5ex}[0pt]{\Large $D^{0}$} &  14         &       &       &       &       &       &       &       &       &       &       &       &        &       &  6.9   & ---   \\
                 &  15         &       &       &       &       &       &       &       &       &       &       &       &        &       &        & ---   \\
\hline \hline
\end{tabular}
}
\vspace*{-0.3cm}
\end{center}
\end{table*}
\begin{table*}[p]
\caption{\label{tab:deltaecuts}Optimized $\Delta E$ selection criteria used for all $B \to D^{(*)} \Dbar^{(*)}$ modes.  Selected events in a given mode must have 
                               $|\Delta E|$ (in \mev) less than the given value.  The $D^{(*)}$ decay modes $1-15$ are defined in Table~\ref{tab:modekey} above.
                               Elements with ``---'' above and on the diagonal are modes that are unused since, due to high 
                               backgrounds, they do not help to increase signal sensitivity.}
\begin{center}
\scalebox{0.83}{
\begin{tabular}{ll|ccccc|ccccc|c|cccc}
\hline \hline
          &                    &  \multicolumn{5}{|c|}{\rule[0mm]{0mm}{5mm}\raisebox{0.5mm}[0mm]{\Large $D^{*-}$}} & 
                                  \multicolumn{5}{|c|}{\rule[0mm]{0mm}{5mm}\raisebox{0.5mm}[0mm]{\Large $\Dstarzb$}} &
                                  \multicolumn{1}{|c|}{\rule[0mm]{0mm}{5mm}\raisebox{0.5mm}[0mm]{\Large $D^{-}$}}  &
                                  \multicolumn{4}{|c}{\rule[0mm]{0mm}{5mm}\raisebox{0.5mm}[0mm]{\Large $\Dzb$}} \\
          &                    &   1   &   2   &   3   &   4   &   5   &   6   &   7   &   8   &   9   &  10   &  11   &    12  &   13  &   14   &   15  \\
\hline
          &          1         & 35.5  & 33.8  & 30.4  & 35.2  & 25.5  & 35.7  & 21.0  & 26.0  &  ---  & 43.6  & 18.0  &  18.1  & 20.2  & 17.1   & 19.0  \\
          &          2         &       & 34.5  & 29.6  & 23.5  & 27.4  & 40.9  & 23.9  & 21.4  &  ---  & 29.3  & 19.4  &  25.9  &  ---  & 19.5   &  ---  \\
{\Large $D^{*+}$}&   3         &       &       & 23.5  & 23.7  & 18.2  & 34.0  & 30.6  & 20.6  &  ---  & 27.3  & 18.6  &  19.0  & 20.4  & 17.1   &  ---  \\
          &          4         &       &       &       &  ---  &  ---  &  ---  &  ---  &  ---  &  ---  &  ---  & 21.9  &  16.9  & 19.7  &  ---   &  ---  \\
          &          5         &       &       &       &       &  ---  &  19.1 &  ---  &  ---  &  ---  &  ---  &  ---  &  16.4  &  ---  &  ---   &  ---  \\
\hline
          &          6         &       &       &       &       &       &  35.1 & 23.0  & 27.3  & 25.5  &  ---  & 23.9  &  17.4  & 19.6  & 17.4   &  ---  \\
          &          7         &       &       &       &       &       &       &  ---  &  ---  &  ---  &  ---  & 20.0  &   ---  &  ---  &  ---   &  ---  \\
{\Large $D^{*0}$}&   8         &       &       &       &       &       &       &       &  ---  &  ---  &  ---  & 16.6  &   ---  &  ---  &  ---   &  ---  \\
          &          9         &       &       &       &       &       &       &       &       &  ---  &  ---  &  ---  &  24.5  &  ---  &  ---   &  ---  \\
                 &  10         &       &       &       &       &       &       &       &       &       &  ---  &  ---  &   ---  &  ---  &  ---   &  ---  \\
\hline
{\Large $D^{+}$} &  11         &       &       &       &       &       &       &       &       &       &       & 15.1  &  15.5  & 19.2  & 15.4   & 15.5  \\
\hline
                 &  12         &       &       &       &       &       &       &       &       &       &       &       &   ---  & 18.7  & 16.1   &  ---  \\
                 &  13         &       &       &       &       &       &       &       &       &       &       &       &        &  ---  & 19.0   &  ---  \\
\raisebox{1.5ex}[0pt]{\Large $D^{0}$} &  14         &       &       &       &       &       &       &       &       &       &       &       &        &       & 15.9   &  ---  \\
                 &  15         &       &       &       &       &       &       &       &       &       &       &       &        &       &        &  ---  \\
\hline \hline
\end{tabular}
}
\end{center}
\end{table*}

\begin{table*}[t!]
\caption{\label{tab:effs}
Elements of the efficiency and crossfeed matrix $\epsilon_{ij}$, and their respective uncertainties, used to calculate the branching fractions and charge asymmetries, as
described in the text.  {\bf\boldmath All values are in units of $10^{-4}$.}  
Uncertainties on the last digit(s) are given in parentheses.  
Elements with ``---'' correspond to values that are zero (to three digits after the decimal point).  
The column corresponds to the generated mode and the row corresponds to the reconstructed mode.
}
\begin{center}
\scalebox{1.0}{
\begin{tabular}{lcccccccccc}
\hline \hline
Mode  &$\Dstarp\Dstarm$ &$\Dstarpm\Dmp$  & $\Dp\Dm$   &  $\Dstarz\Dstarzb$   & $\Dstarz\Dzb$   & $\Dz\Dzb$   & $\Dstarp\Dstarzb$   & $\Dstarp\Dzb$  & $\Dp\Dstarzb$   &  $\Dp\Dzb$ \\
\hline
$\Dstarp\Dstarm$   & 14.24(6) & 0.010(3) &  ---  &  ---  &  ---  &  ---  & 0.18(1) & ---   &  ---  &   --- \\
$\Dstarpm\Dmp$     & 0.020(3) & 11.52(6) &  ---  &  ---  &  ---  &  ---  & 0.010(3) & 0.040(3) & 0.08(1) &   --- \\
$\Dp\Dm$           &   ---  &   --- & 9.51(8) &  ---  &  ---  &   --- & ---   & ---   &  ---  & 0.010(3) \\
$\Dstarz\Dstarzb$  & 0.080(3) & ---   & ---   & 2.60(2) & 0.030(3) & ---   & 0.42(1) & 0.010(3) &  ---  &   --- \\
$\Dstarz\Dzb$      &   --- &   --- &   --- & 0.020(3) & 3.40(2) &  ---  & 0.010(3) & 0.46(1) & 0.010(3) &   --- \\
$\Dz\Dzb$          &   --- &   --- &   --- &  ---  & 0.010(3) & 12.02(10) &  ---  & 0.010(3) & 0.020(3) &  ---  \\
$\Dstarp\Dstarzb$  & 2.60(2) & ---   & ---   & 0.23(1) & 0.010(3) &   --- & 7.52(4) & 0.07(1) &  ---  &   --- \\
$\Dstarp\Dzb$      & 0.040(3) & 0.06(2) & ---   &  ---  & 0.11(5) &   --- & 0.03(2) & 13.51(25) & 0.040(3) &  ---  \\
$\Dp\Dstarzb$      &   --- & 0.41(1) &   --- & 0.010(3) & 0.010(3) &  ---  &  ---  & 0.070(3) & 3.70(3) &  --- \\
$\Dp\Dzb$          &   --- & 0.020(3) & 0.06(1) &  ---  &  ---  & 0.050(3) &  ---  & 0.010(3) & 0.020(3) & 14.93(9) \\
\hline \hline
\end{tabular}
}
\end{center}
\end{table*}

\section{Efficiency and crossfeed determination}\label{sec:effs}

The efficiencies are determined using fits to \mes distributions of signal MC events that pass all selection criteria in $\masslik$, $|\Delta E|$, 
$\mathcal{F}$, and $L$.
There is a small, but non-negligible probability that a signal $B$ decay
of mode $i$ is reconstructed as a different signal decay mode $j$.
We refer to this as crossfeed.
Thus, efficiencies can be represented as a matrix $\epsilon_{ij}$.  
where each contributing generated event is weighted by the $D$ and $\Dbar$ decay mode branching fractions. 
To determine the elements of $\epsilon_{ij}$, we fit the \mes distributions of signal MC events generated as $B$ decay mode $i$
and reconstructed as $B$ decay mode $j$.  The distributions are modeled as the sum of signal and background probability distribution functions (PDFs), where the PDF for 
the signal is a Gaussian distribution centered around the $B$ mass, and the PDF for background is an 
empirical function~\cite{argus} of the form
\begin{equation}
\label{eq:argus}
f(x) \propto x\sqrt{1-x^2}\exp[-\kappa(1-x^2)],
\end{equation}
where we define $x \equiv 2\mes/\sqrt{s}$, and $\kappa$ is a parameter determined by the fit.  In \BB MC samples containing signal and background decays, 
we find that the \mes distribution is well-described 
by adding a simple Gaussian function to the empirical shape in Eq.~\ref{eq:argus}.  We fit the \mes distributions of signal MC events generated as mode $i$ and passing
selection criteria in mode $j$ to the above distribution
by minimizing the $\chi^2_{ij}$ of each fit with respect to $\kappa_{ij}$ (the $\kappa$ parameter for each mode $(i,j)$), the number of signal events $N^s_{ij}$, 
and the number of background 
events $N^b_{ij}$.
We determine the efficiencies $\epsilon_{ij}$ as $N^s_{ij}/N^g_i$, where $N^g_i$ is the total number of signal MC events that were generated in mode $i$.
The diagonal elements of the $\epsilon_{ij}$ matrix (\textit{i.e.} the numbers typically denoted as ``efficiencies'') are in the range (0.2 -- 1.5)$ \times 
10^{-3}$.
The main crossfeed source is misidentification between $D^{*0}$ and $D^{*\pm}$ candidates.  
The matrix $\epsilon_{ij}$ and the uncertainties on the elements of this matrix are given in Table~\ref{tab:effs}.
Crossfeed between different $D$ submodes (\textit{i.e.} mode numbers 12--15 in Table~\ref{tab:modekey}) is negligible. 

\section{Branching fraction results}\label{sec:bfs}

In order to determine the number of signal events in each mode, one must not only account for background which is distributed according to
combinatorial phase space, but also for background which can have a different distribution in \mes.
It is possible for a component of the background to have an \mes distribution
with a PDF that is more similar to signal (\textit{i.e.} a Gaussian distribution centered around 
the $B$ mass) than to a phase-space distribution.  Such a component
is known as ``peaking'' background and typically derives from background events that have the 
same or similar final state particles as the signal decay mode.  
For example, in \Bztodd, peaking background primarily comes from the decays $\Bz \rightarrow DKX$ or
$\Bz \rightarrow D\pi X$, where $D \rightarrow \Kpipi$ and $X$ is $K^0$,
$\rho$, $a_1$ or $\omega$, and
the light mesons ($KX$) or ($\pi X$) fake a $D \rightarrow \Kpipi$ decay.
The optimization procedure that was detailed in Sec.~\ref{sec:analopti_sele} eliminates decay submodes that have a 
large enough amount of peaking (in addition to combinatorial) background to decrease, rather than increase, the sensitivity for a particular decay;
the final selection was detailed in Tables~\ref{tab:globalcuts}, \ref{tab:masslikcuts}, and~\ref{tab:deltaecuts}.
We determine the
amount of peaking background $P_i$ in each $B$ decay mode $i$ via fitting the \mes distributions of \BB MC simulated events. 
We minimize the $\chi^2_{i}$ of each fit, allowing the variables
$\kappa^P_{i}$ (representing the ``ARGUS parameter'' described earlier), the number of expected peaking background events in data $P_{i}$, and the number of 
phase-space background events $N^{\rm MC bkg}_{i}$, to float.  The fitted number of peaking background events $P_i$ 
is compatible with zero, within two standard deviations, for all 
modes $i$.

\begin{figure*}[p]
\begin{center}
\scalebox{0.52}{\includegraphics{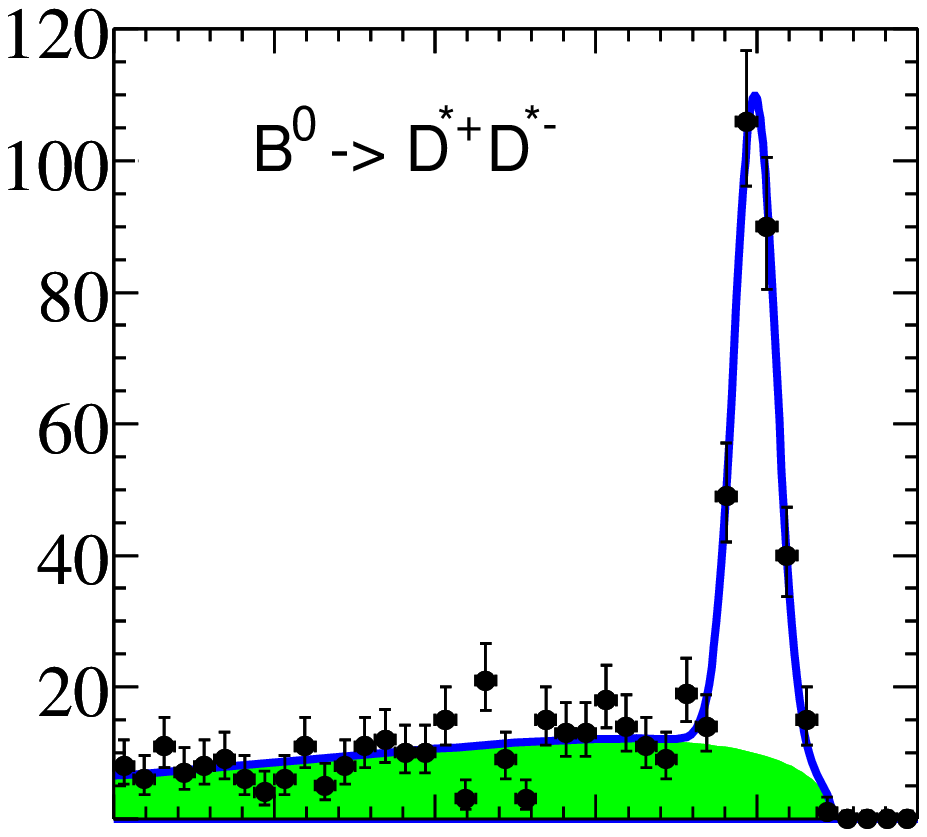}}
\hspace*{0.5cm}
\scalebox{0.52}{\includegraphics{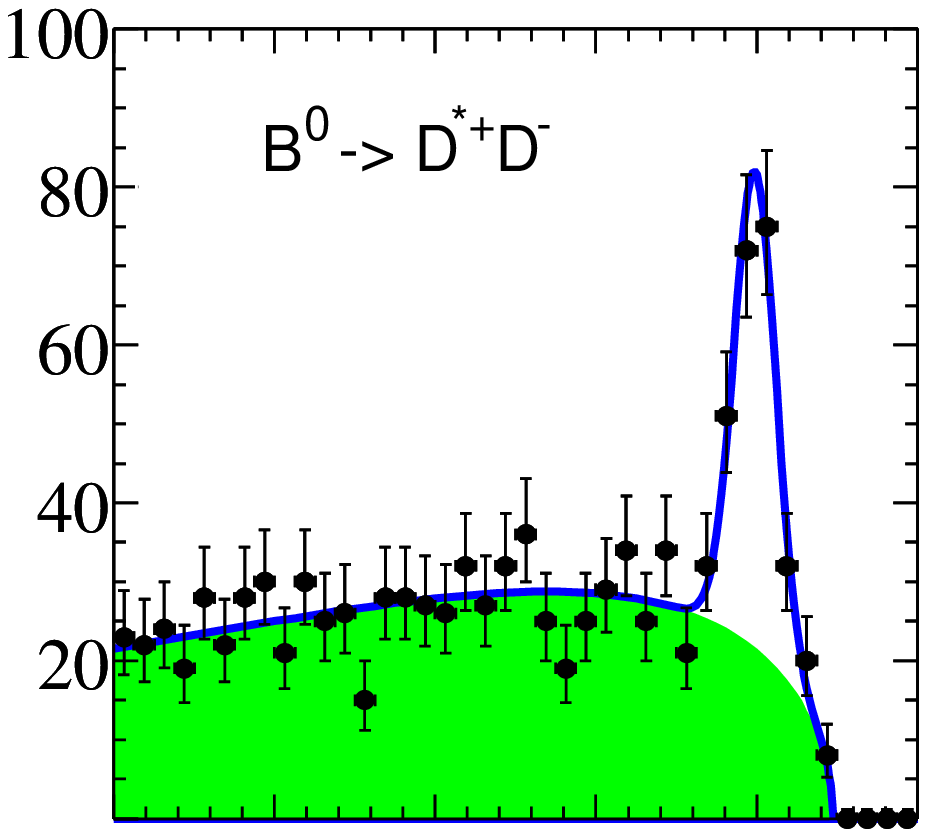}}

\vspace*{-0.2cm}

\hspace*{0.04cm}
\scalebox{0.52}{\includegraphics{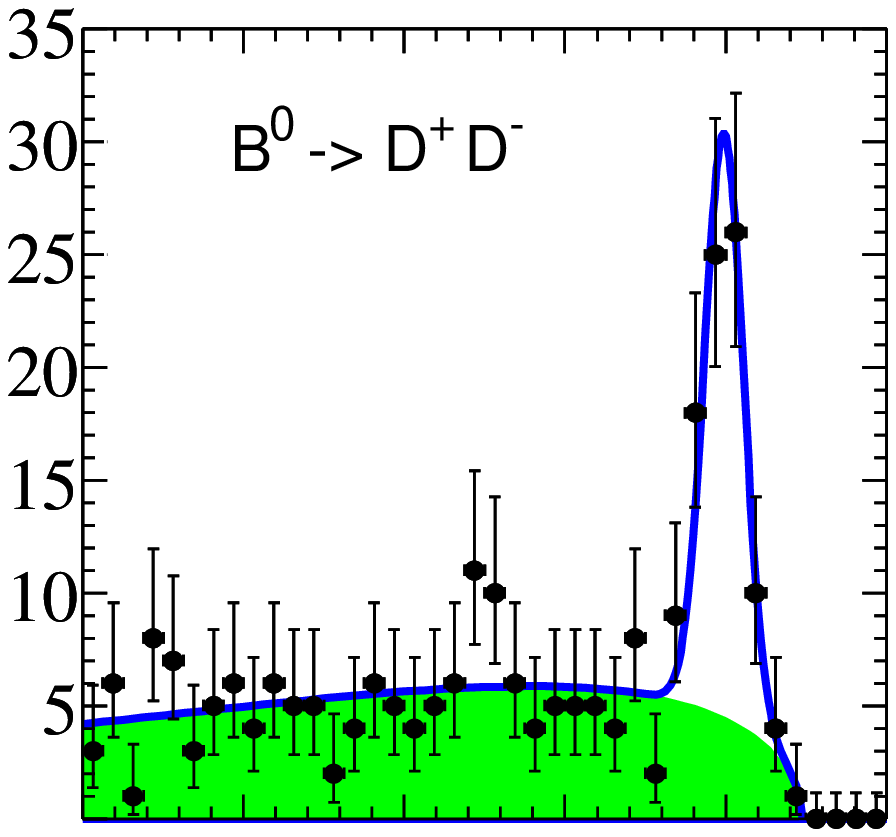}}
\hspace*{0.65cm}
\scalebox{0.52}{\includegraphics{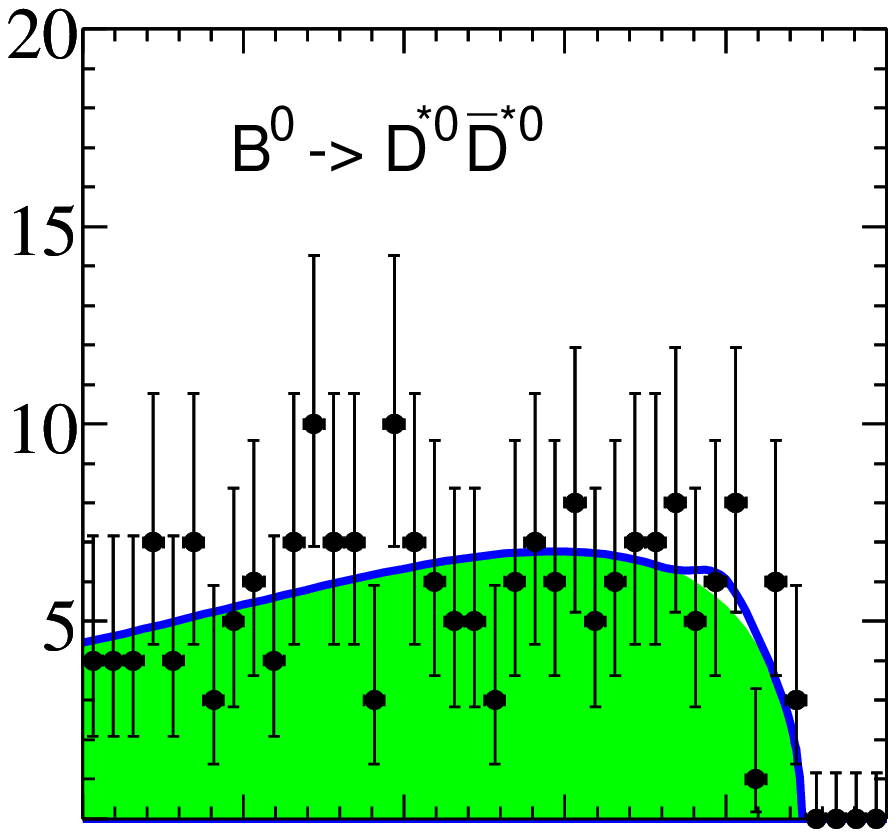}}

\vspace*{-0.2cm}

\hspace*{-0.49cm}
\scalebox{0.52}{\includegraphics{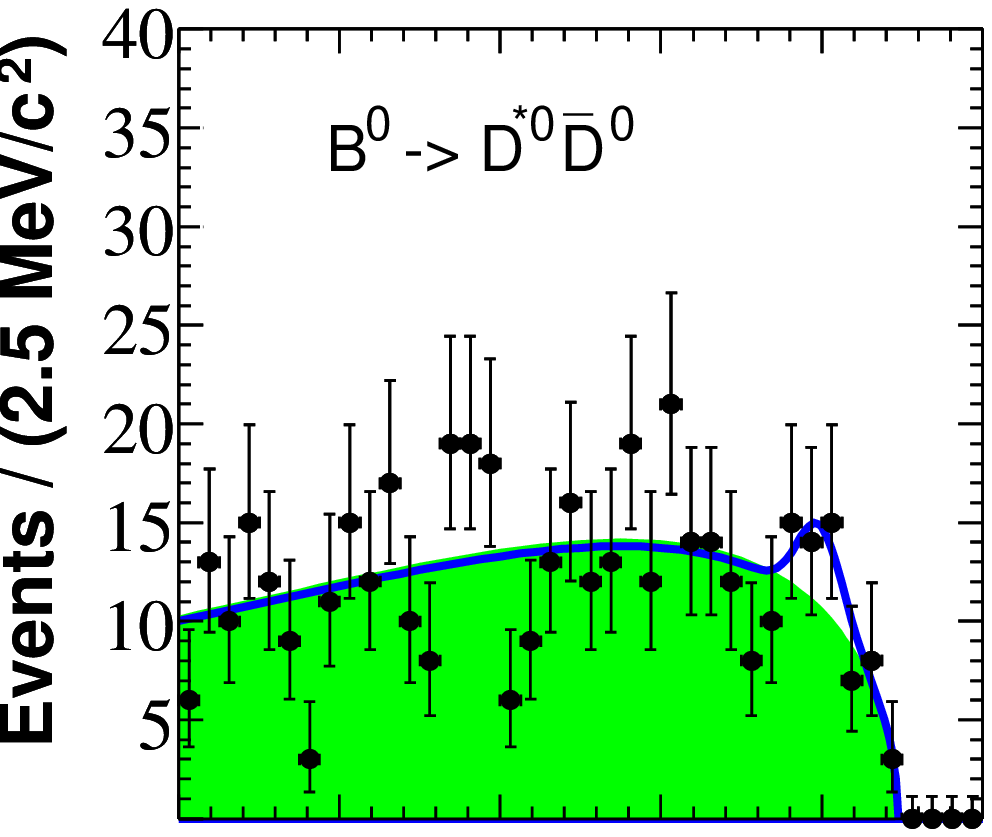}}
\hspace*{0.65cm}
\scalebox{0.52}{\includegraphics{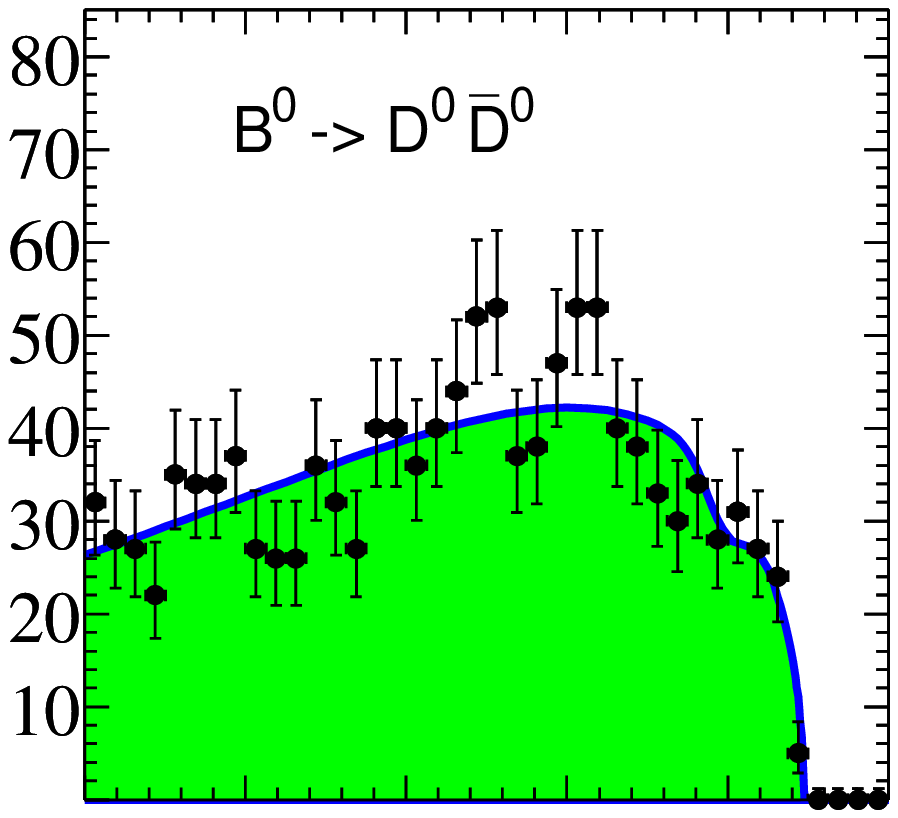}}

\vspace*{-0.2cm}

\hspace*{-0.12cm}
\scalebox{0.52}{\includegraphics{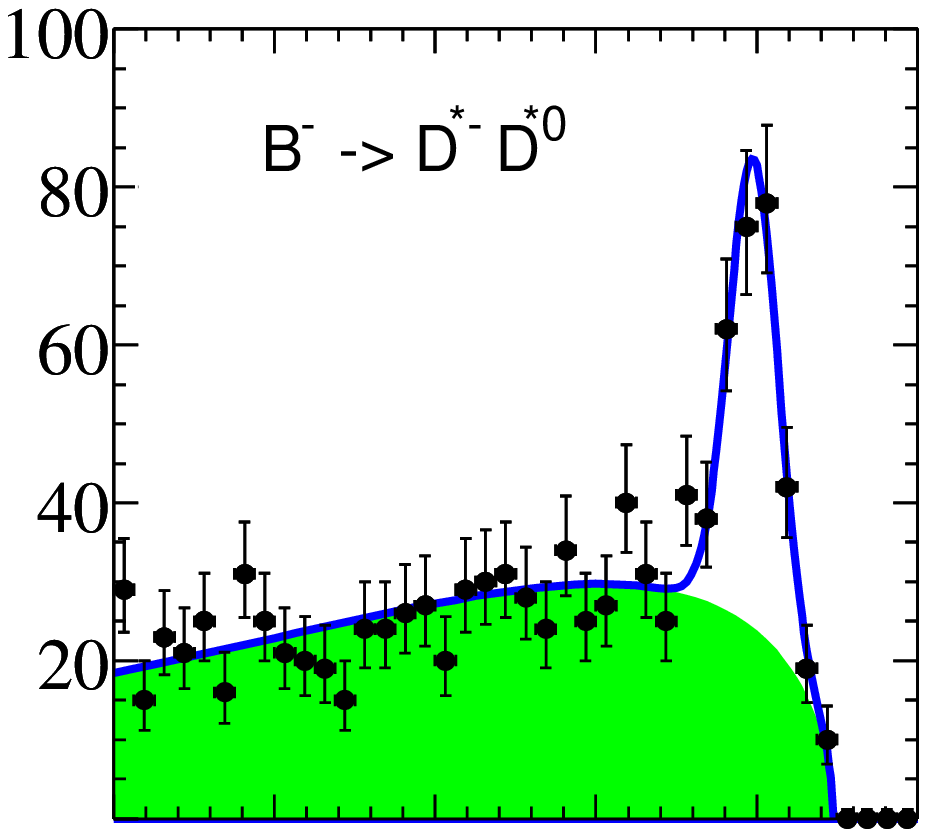}}
\hspace*{0.65cm}
\scalebox{0.52}{\includegraphics{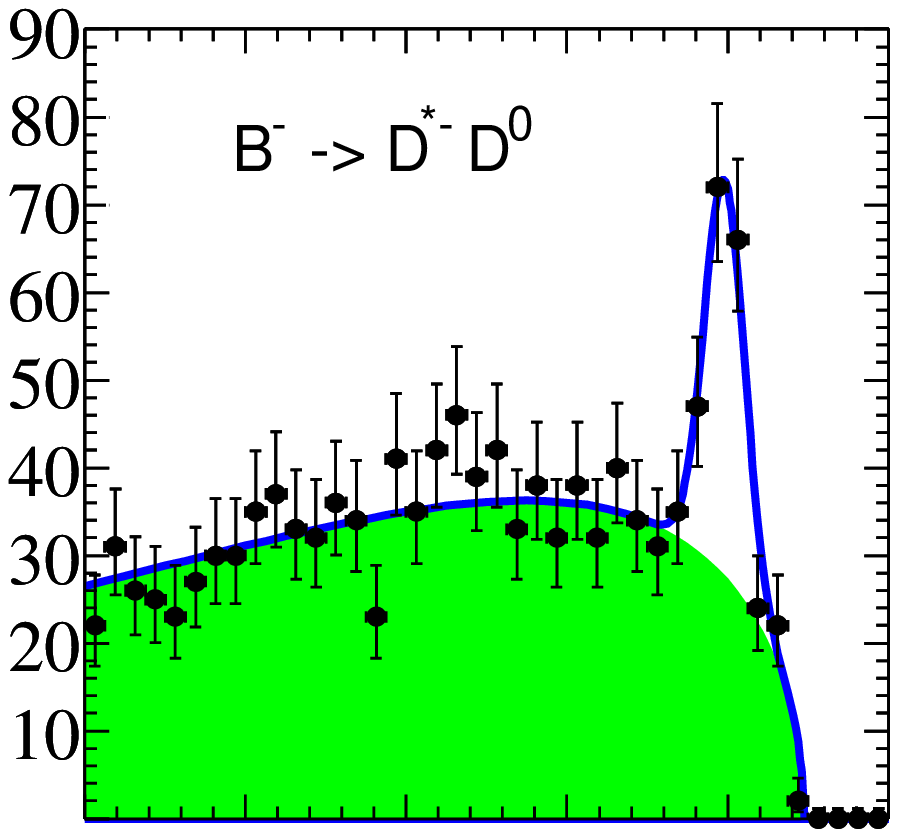}}

\vspace*{-0.2cm}

\hspace*{0.25cm}
\scalebox{0.52}{\includegraphics{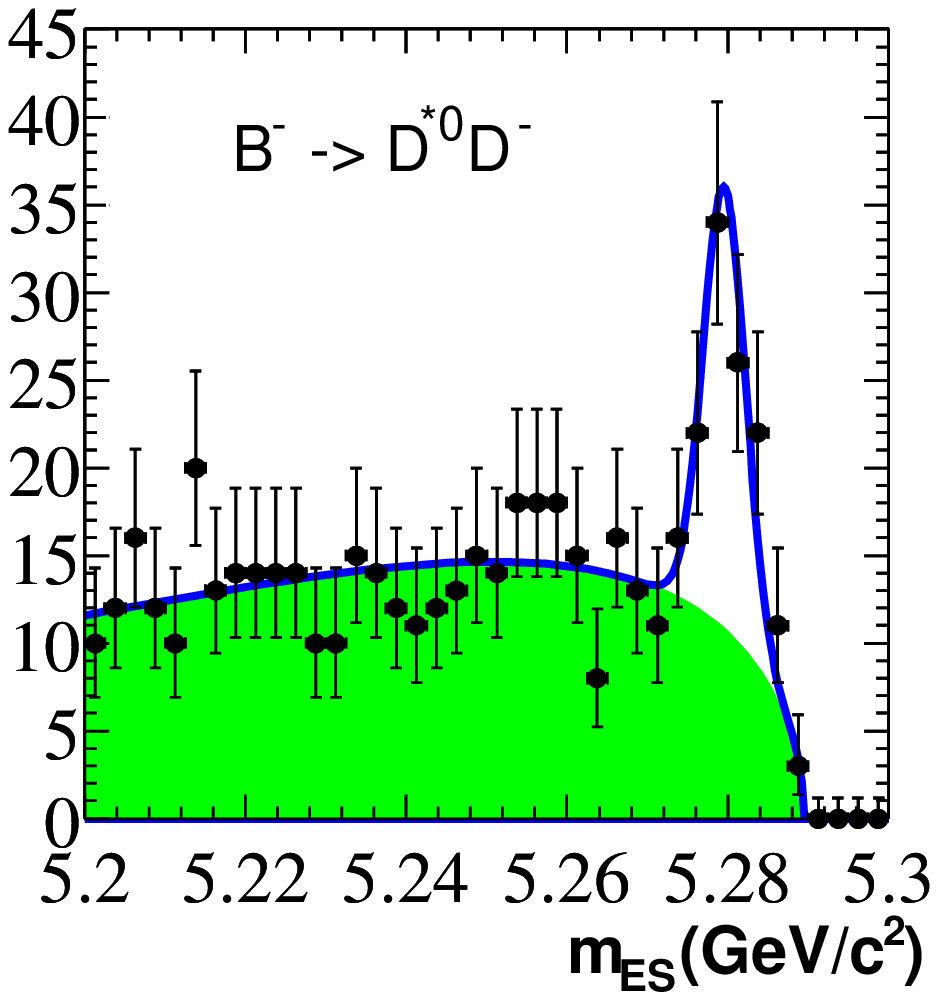}}
\hspace*{0.32cm}
\scalebox{0.52}{\includegraphics{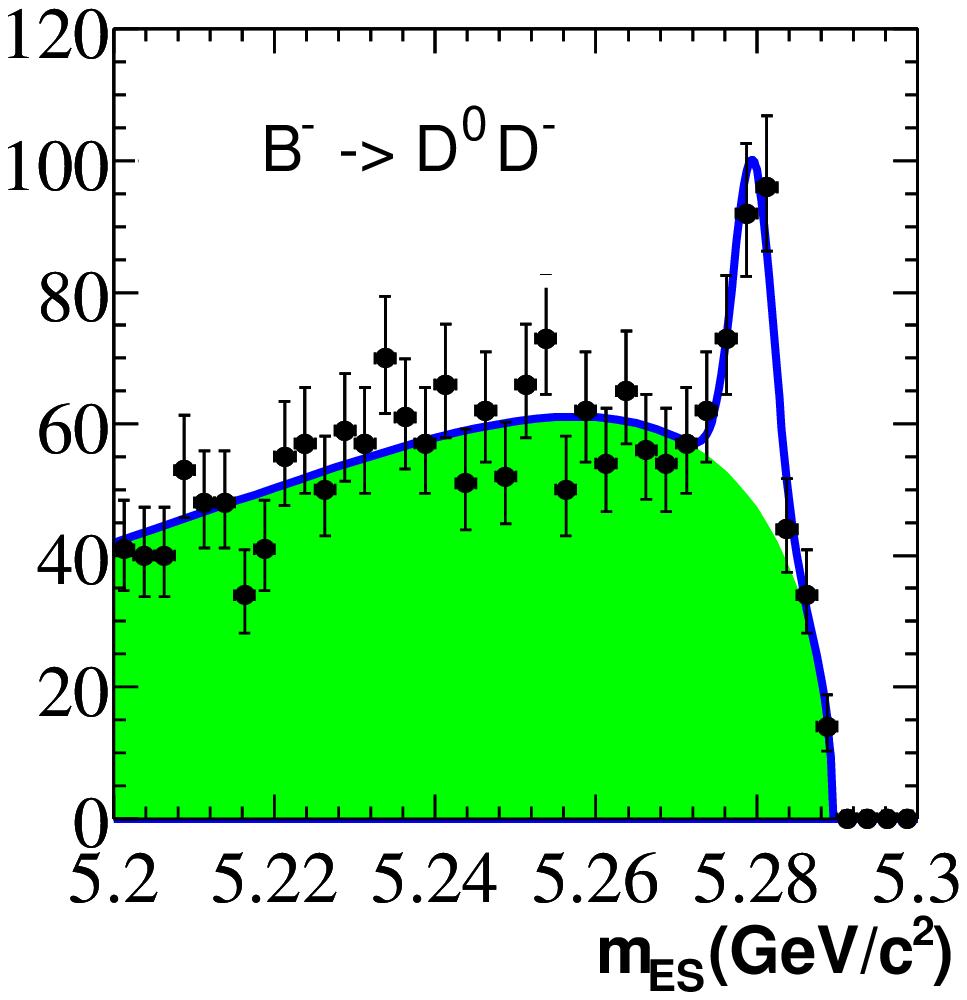}}

\vspace*{-0.2cm}
\end{center}
\caption{
\label{fig:data_dstdst}
Distributions of \mes for selected candidates in each $D^{(*)}\Dbar^{(*)}$ mode.
The error bars represent the statistical errors only.
The solid lines represent the fits to the data, and the shaded areas
the fitted background.
}
\end{figure*}

\begin{table*}[t]
\caption{\label{tab:bfsacps}
Results of the fits for the ten signal decay modes: the number of events for fitted signal $N^{\rm sig}$,
the peaking background $P$, and the crossfeed $C$, the branching fractions
$\mathcal{B}$, 90\% C.L.~upper limits on
branching fractions, previous measurements of branching fractions
(for modes that have previous measurements), and charge asymmetries.  The uncertainties are statistical.  For
the final branching fraction and charge asymmetry results,
the systematic errors are also given.
}
\begin{center}
\begin{tabular}{lr@{$\pm$}lr@{$\pm$}lr@{$\pm$}lr@{}c@{}lccc}
\hline\hline
\textbf{Mode}                            & \multicolumn{2}{c}{$N^{\rm sig}$} 
                                                         & \multicolumn{2}{c}{$P$}  & \multicolumn{2}{c}{$C$}   
                                                                           &  \multicolumn{3}{c}{{\boldmath $\mathcal{B}$} $(10^{-4})$}        &  
                                                                   $\begin{array}{c} \mbox{\textbf{U.L.}} \\ (10^{-4}) \end{array}$  & 
                                                                   $\begin{array}{c} \mbox{Previous $\mathcal{B}$} \\ \mbox{results $(10^{-4})$} \end{array}$ 
                                   & {\boldmath $\mathcal{A}_{\CP}$} \\
\hline
{\boldmath \Bz $\to D^{*+}D^{*-}$}       & 270 & 19  & $-1$ &  2 &  4  &  1  &  {\boldmath \BRbztodstdstCV}   & {\boldmath \BRbztodstdstStat}   & {\boldmath \BRbztodstdstSyst}   &                      
& 
      $\begin{array}{c} 8.1 \pm 0.8 \pm 1.1\;[21] \\ 8.3 \pm 1.6 \pm 1.2\;[22] \\ 9.9^{+4.2}_{-3.3} \pm 1.2\;[23] \end{array}$                            & \\
\hline
{\boldmath \Bz $\to D^{*\pm}D^{\mp}$}    & 156 & 17  &  1 &  3 &  2  &  1  &  {\boldmath \BRbztodstdCV}     & {\boldmath \BRbztodstdStat}     & {\boldmath \BRbztodstdSyst}     
&                      &  
      $\begin{array}{c} 8.8 \pm 1.0 \pm 1.3\;[24] \\ 11.7 \pm 2.6^{+2.2}_{-2.5}\;[25] \\ 6.7^{+2.0}_{-1.7} \pm 1.1\;[26] \end{array}$  & 
                                                     {\boldmath $\quad\ACPbztodstdNum\;\mbox{[27]}$} \\
\hline
{\boldmath \Bz $\to D^{+}D^{-}$}         &  63 &  9  &  1 &  2 &  0  &  0  &  {\boldmath \BRbztoddCV}       & {\boldmath \BRbztoddStat}       & {\boldmath \BRbztoddSyst}       &   & 
$1.91 \pm 0.51 \pm 0.30\;[28]$   & \\
\hline
{\boldmath \Bz $\to D^{*0}\Dstarzb$} &  0 &  6  & $-2$ &  2 &  0  &  0  &  \BRbztodstzdstzCV & \BRbztodstzdstzStat & \BRbztodstzdstzSyst & {\boldmath \LIMbztodstzdstzMant} & 
$< 270$~\cite{Aleph1} & \\
\hline
{\boldmath \Bz $\to D^{*0}\Dzb$}  & 10 &  8  & $-2$ &  3 &  1  &  1  &  \BRbztodstzdzCV   & \BRbztodstzdzStat   & \BRbztodstzdzSyst   & {\boldmath \LIMbztodstzdzMant}   &                  
& \\
\hline
{\boldmath \Bz $\to D^{0}\Dzb$}   & $-11$ & 12  & $-8$ & 4 &  0  &  0  &  \BRbztodzdzCV     & \BRbztodzdzStat     & \BRbztodzdzSyst     & {\boldmath \LIMbztodzdzMant}     &                  
& \\
\hline\hline
{\boldmath \Bp $\to D^{*+}\Dstarzb$} & 185 & 20  & $-5$ & 4 & 34  &  4  &  {\boldmath \BRbptodstdstzCV}  & {\boldmath \BRbptodstdstzStat}  & {\boldmath \BRbptodstdstzSyst}  &                      
&  $\begin{array}{c} 10.5^{+3.3}_{-2.8} \pm 2.0\;[26] \\ < 110\;[29] \end{array}$ & 
                                             {\boldmath $\ACPbptodstdstzNum$ } \\
\hline
{\boldmath \Bp $\to D^{*+}\Dzb$}  & 115 & 16  &  1 & 4 &  3  &  1  &  {\boldmath \BRbptodstdzCV}    & {\boldmath \BRbptodstdzStat}    & {\boldmath \BRbptodstdzSyst} & & 
$\begin{array}{c} 4.57 \pm 0.71 \pm 0.56\;[28] \\ < 130\;[29,30] \end{array}$ & 
                                             {\boldmath $\ACPbptodstdzNum$ }   \\
\hline
{\boldmath \Bp $\to D^{+}\Dstarzb$}  &  63 & 11  & 3 &  3 &  9  &  2  &  {\boldmath \BRbptoddstzCV}    & {\boldmath \BRbptoddstzStat}    & {\boldmath \BRbptoddstzSyst} & & 
$< 130$~\cite{Aleph1,Aleph1N} & 
                                             {\boldmath $\ACPbptoddstzNum$ }   \\
\hline
{\boldmath \Bp $\to D^{+}\Dzb$}   & 129 & 20  & $-2$ &  5 &  1  &  1  &  {\boldmath \BRbptoddzCV}      & {\boldmath \BRbptoddzStat}      
& {\boldmath \BRbptoddzSyst}      
& & 
 $\begin{array}{c} 4.83 \pm 0.78 \pm 0.58\;[28] \\ < 67\;[29] \end{array}$  & 
                                             {\boldmath $\ACPbptoddzNum$ }    \\
\hline\hline
\end{tabular}
\end{center}  
\end{table*}

We then fit the actual data to determine the number of reconstructed signal events in each mode.
We fit the \mes distributions of reconstructed $B$ decays that pass all selection criteria in each mode $i$ to a sum of a Gaussian distribution and a phase space
distribution (Eq.~\ref{eq:argus}), similar to the PDFs used for efficiency and peaking background fits described above.  We minimize the
$\chi^2_{i}$ of each data fit, allowing the parameter $\kappa_{i}$, the number of signal events in data $N^{\rm sig}_{i}$, and the number of background events in data 
$N^{\rm bkg}_{i}$, each to float. 
The \mes distributions and the results of the fits are shown in Fig.~\ref{fig:data_dstdst}.
The branching fractions $\mathcal{B}_i$ are then determined via the equation
\begin{equation}
\label{eq:bfs}
\sum_j \epsilon_{ij} \mathcal{B}_j N_B = N^{\rm sig}_i - P_i
\end{equation}
where $N_B = N_{\BB} = (231.7 \pm 2.6) \times 10^{6}$ is the total number of charged or neutral $B$ decays in the data sample,
assuming equal production rates of charged and neutral $B$ pairs.

We determine the branching fractions as
\begin{equation}
\mathcal{B}_i = \sum_j\epsilon^{-1}_{ij}(N^{\rm sig}_j - P_j)/N_B,
\end{equation}
(where $\epsilon^{-1}_{ij}$ is the inverse of matrix $\epsilon_{ij}$) yields
the branching fractions given in Table~\ref{tab:bfsacps}.  Note that the measured branching fractions for the three modes 
$\Bz \to D^{(*)0}\Dbar^{(*)0}$ are not significantly greater than zero.  Thus, we have determined upper limits on
the branching fractions for these modes.  The 90\% confidence level (C.L.) upper limits quoted in Table~\ref{tab:bfsacps} are 
determined using the Feldman-Cousins method~\cite{FeldmanCousins} and include all systematic uncertainties detailed below. 
Since the branching fractions can be correlated through the use of Eq.~\ref{eq:bfs}, we also provide the covariance matrix, with 
all systematic uncertainties included, in Table~\ref{tab:covBF}. The covariance matrix is obtained via the approximation given in~\cite{matinverr}.

\begin{table*}[t]
\caption{Covariances of $B \to D^{(*)}\bar{D}^{(*)}$ branching fractions (with all systematic uncertainties included), in units of $10^{-8}$.}
\begin{center}
\begin{tabular}{rcccccccccc}
\hline \hline
Mode  &$\Dstarp\Dstarm$ &$\Dstarpm\Dmp$  & $\Dp\Dm$   &  $\Dstarz\Dstarzb$   & $\Dstarz\Dzb$   & $\Dz\Dzb$   & $\Dstarp\Dstarzb$   & $\Dstarp\Dzb$  & $\Dp\Dstarzb$   &  $\Dp\Dzb$ \\
\hline		                                                                	          
$\Dstarp\Dstarm$  & 1.26 &    0.55 &    0.22 & $-$0.15 &    0.07 & $-$0.01 &    0.73 &    0.33 &    0.54 &    0.30 \\
$\Dstarpm\Dmp$    &      &    0.91 &    0.26 & $-$0.08 &    0.04 & $-$0.01 &    0.46 &    0.19 &    0.37 &    0.26 \\
$\Dp\Dm$          &      &         &    0.39 & $-$0.03 &    0.02 &    0.00 &    0.16 &    0.08 &    0.26 &    0.16 \\
$\Dstarz\Dstarzb$ &      &         &         &    1.27 & $-$0.04 &    0.00 & $-$0.53 & $-$0.06 & $-$0.13 & $-$0.05 \\
$\Dstarz\Dzb$     &      &         &         &         &    1.25 &    0.00 &    0.07 & $-$0.02 &    0.05 &    0.02 \\
$\Dz\Dzb$         &      &         &         &         &         &    0.22 & $-$0.01 &    0.00 & $-$0.01 &    0.00 \\
$\Dstarp\Dstarzb$ &      &         &         &         &         &         &    2.60 &    0.31 &    0.55 &    0.27 \\
$\Dstarp\Dzb$     &      &         &         &         &         &         &         &    0.43 &    0.19 &    0.11 \\
$\Dp\Dstarzb$     &      &         &         &         &         &         &         &         &    2.61 &    0.27 \\
$\Dp\Dzb$         &      &         &         &         &         &         &         &         &         &    0.53 \\
\hline \hline
\end{tabular}
\label{tab:covBF}
\end{center}
\end{table*}

\section{Branching fraction systematic uncertainties}\label{sec:bfsysts}

\begin{table*}[t]
\caption{Estimates of branching fraction systematic uncertainties (as percentages of the absolute values of the branching fraction central values) for all $B$ modes, after propagating the errors through Eq.~\ref{eq:bfs}.
The totals are the sums in quadrature of the uncertainties in each column.
Note that the term ``Dalitz weight'' refers to the selection on the reconstructed invariant masses of the $\Km\pip$ and $\Km\piz$ pairs
for $\Dz \to \Km\pip\piz$ decays that was described in Sec.~\ref{sec:dmeson_sele}.
}
\begin{center}
\begin{small}
\begin{tabular}{lccccccccccc}
\hline \hline
Mode  & $\Dstarp\Dstarm$ & $\Dstarpm\Dmp$  & $\Dp\Dm$   &  $\Dstarz\Dstarzb$   & $\Dstarz\Dzb$   & $\Dz\Dzb$   & $\Dstarp\Dstarzb$   & $\Dstarp\Dzb$  & $\Dp\Dstarzb$   &  $\Dp\Dzb$ \\
\hline
$D^{*+}$ BFs        & 1.4 & 0.7 & 0.0 & 0.9 & 0.1 & 0.0 & 0.7 & 0.7 & 0.0 & 0.0 \\
$D^{*0}$ BFs        & 0.0 & 0.0 & 0.0 & 4.9 & 1.6 & 0.0 & 2.1 & 0.0 & 4.4 & 0.0 \\
$D^{0}$ BFs         & 5.0 & 2.7 & 0.0 & 7.4 & 3.7 & 5.7 & 5.2 & 4.5 & 3.3 & 2.7 \\
$D^{+}$ BFs        & 1.4 & 6.5 & 13.2 & 0.1 & 0.2 & 0.4 & 0.1 & 0.3 & 6.5 & 6.5 \\
Tracking efficiency & 7.9 & 6.5 & 4.8 & 7.9 & 3.0 & 4.7 & 6.0 & 6.0 & 3.8 & 4.4 \\
\KS efficiency      & 0.3 & 0.2 & 0.0 & 0.0 & 0.1 & 0.0 & 0.1 & 0.2 & 0.3 & 0.2 \\
Neutrals efficiency & 2.5 & 1.0 & 0.0 & 8.4 & 2.9 & 1.9 & 4.6 & 1.6 & 4.3 & 1.0 \\
Kaon identification & 4.6 & 4.7 & 5.0 & 7.3 & 4.9 & 5.4 & 5.0 & 4.6 & 4.6 & 4.7 \\
\masslik cut        & 1.1 & 1.1 & 1.1 & 1.1 & 1.1 & 1.1 & 1.1 & 1.1 & 1.1 & 1.1 \\
$\cal{F}$ cut       & 0.0 & 0.0 & 0.9 & 0.9 & 0.9 & 0.9 & 0.9 & 0.9 & 0.9 & 0.9 \\
$L$ cut             & 0.0 & 0.0 & 0.8 & 0.8 & 0.8 & 0.8 & 0.0 & 0.8 & 0.8 & 0.8 \\
\DeltaE cut         & 1.1 & 1.1 & 1.1 & 1.1 & 1.1 & 1.1 & 1.1 & 1.1 & 1.1 & 1.1 \\
Dalitz weight cut   & 1.0 & 0.5 & 0.0 & 1.4 & 0.2 & 1.0 & 1.0 & 0.8 & 0.7 & 0.5 \\
P($\chi^2$) cut     & 3.8 & 3.8 & 3.8 & 3.8 & 3.8 & 3.8 & 3.8 & 3.8 & 3.8 & 3.8 \\
Fit model          & 1.8 & 3.6 & 3.1 & 5.4 & 6.7 & 44.6 & 4.9 & 2.8 & 7.0 & 3.6 \\
Spin alignment      & 1.0 & 0.0 & 0.0 & 6.1 & 0.0 & 0.1 & 4.1 & 0.0 & 0.0 & 0.0 \\
Peaking background & 0.9 & 2.0 & 2.9 & 24.5 & 32.3 & 144.6 & 3.1 & 3.4 & 4.9 & 4.0 \\
Crossfeed           & 0.4 & 0.6 & 0.8 & 1.9 & 1.1 & 1.6 & 0.6 & 0.4 & 1.0 & 0.6 \\
$N_{\BB}$           & 1.1 & 1.1 & 1.1 & 1.1 & 1.1 & 1.1 & 1.1 & 1.1 & 1.1 & 1.1 \\
\hline
  Total  & 12.0 & 12.3 & 16.1 & 31.0 & 34.2 & 151.7 & 13.6 & 11.0 & 14.8 & 11.9 \\
\hline \hline
\end{tabular}
\end{small}
\label{tab:brsysts}
\end{center}
\end{table*}
Table~\ref{tab:brsysts} shows the results of our evaluation of the systematic uncertainties on the branching fraction measurements.
\paragraph{Submode branching fractions} The central values and uncertainties on the branching fractions of the $D$ and \Dstar mesons are propagated into the calculation
of the branching fraction measurements.  The world average measurements~\cite{PDG2004} are used.
\paragraph{Charged track finding efficiency} From studies of absolute tracking efficiency, we assign a systematic uncertainty of 0.8\% 
per charged track on the efficiency of finding tracks 
other than slow pions from charged \Dstar decays and daughters of \KS decays.  For the slow pions, we assign a systematic uncertainty of 2.2\% each, 
as determined from a separate efficiency 
study (using extrapolation of slow tracks found in the SVT into the DCH tracking detector and vice-versa).
Track finding efficiency uncertainties are treated as 100\% correlated among the tracks in a candidate.
These uncertainties are weighted by the $D$ and \Dstar branching fractions.
\paragraph{\KS reconstruction efficiency} From a study of the \KS reconstruction efficiency (using an inclusive data sample of events containing one or more \KS, as well as 
corresponding MC samples), we assign a 2.5\% 
systematic uncertainty for all modes containing a \KS.
The value 2.5\% comes from the statistical uncertainty in the ratio of data to MC yields and the variation of this ratio over different selection criteria.
The uncertainty is weighted by the $D$ and \Dstar branching fractions.
\paragraph{\piz and $\gamma$ finding efficiency} From studies of the neutral particle finding efficiency through the ratios of $\tau^+ \to \rho^+(\pi^+\pi^0)\nu$ to $\tau^+ \to \pi^+\nu$ between data and MC, we assign a 3\% systematic uncertainty per \piz, including 
the slow \piz from \Dstar and \Dstarz decays.  For isolated photons from \Dstarz decays, we assign a 1.8\% systematic uncertainty, 100\% correlated with the \piz efficiency uncertainty.
These uncertainties are weighted by the $D$ and \Dstar branching fractions.
\paragraph{Charged kaon identification} We assign a systematic uncertainty of 2.5\% per charged kaon, according to a study of kaon particle identification efficiency
(using kinematically-reconstructed $\Dz \to \Km\pip$ candidates).
The uncertainty is weighted by the $D$ and \Dstar branching fractions.
\paragraph{Other selection differences between data and MC} Differences in momentum measurement, decay vertex finding efficiency, etc., can result in additional differences between 
efficiencies
in data and in MC.  We use a sample of the more abundant $B^0 \to D_s^{*+}\Dstarm$ events in data, selected in a similar manner as the
$B \to D^{(*)}\Dbar^{(*)}$ modes, to determine these uncertainties.
To estimate the systematic error arising from differences
between the data and MC $D$ and $D^*$ mass resolutions, we
calculate the number of $D_s^*D^*$ events seen in the data and MC
as a function of the \masslik cut, while fixing the other selection criteria to their nominal values. The number of observed events is
extracted from a fit to the \mes distribution. We then plot the ratio of the data yield
($N_{\rm data}$) to the MC yield ($N_{\rm MC}$) as a function of the
\masslik cut over a range of values that gives the same efficiencies as
in the $D^{(*)}\Dbar^{(*)}$ analyses.  We find the rms of the
$N_{\rm data}/N_{\rm MC}$ ratio and assign this as a systematic
uncertainty for applying this cut.  The same technique is used to
determine the systematic uncertainties from all other selection criteria in Table~\ref{tab:brsysts}: the selections on 
$\cal{F}$, $L$, \DeltaE, the reconstructed invariant masses for $\Dz \to \Km\pi+\piz$ (``Dalitz weight''), and vertex P($\chi^2$).
\paragraph{Fit model} The data yield is obtained from an \mes fit where the mean ($\mu$) and
width ($\sigma$) of the $B$ mass and the end-point ($\sqrt{s}/2$) of the phase-space distribution (Eq.~\ref{eq:argus}) are fixed. These parameters are
estimated and have associated uncertainties.  The nominal value of $\sigma$ is determined from signal MC for each $B$ decay mode.
To estimate the systematic uncertainty due to possible differences
between the \mes resolutions in data and signal MC, we first look at this
difference ($\Delta\sigma = \sigma_{\rm data} - \sigma_{\rm MC}$) for those
modes with high purity, including our control sample.
These differences are consistent with zero, justifying our use of $\sigma_{\rm MC}$ in obtaining the data
yield. We then find the weighted average of $\Delta\sigma$, which is
given by $(0.11 \pm 0.08)$ \mevcc.
As a conservative estimate, we repeat the data yield determinations
by moving $\sigma$ up and down by 0.2 \mevcc, and take the average of the absolute values of the
changes in each data yield as the systematic uncertainty of
fixing $\sigma$ to the MC value for that $B$ mode.
A combined fit of common modes in data is used to determine the nominal values for $\mu$ and for the endpoint of the \mes distribution $\sqrt{s}/2$.
Hence, we move the parameters up and down by their fitted errors
(0.2 \mevcc for $\mu$ and 0.1 \mevcc for $\sqrt{s}/2$) to obtain their
corresponding systematic uncertainties.
The quadratic sum of the three uncertainties from $\mu$, $\sigma$ and
$\sqrt{s}/2$ gives the systematic uncertainty of the fit model for each $B$ mode.
\paragraph{Spin-alignment dependence} The $\Bztodstdst$, $\Bztodstzdstz$, and $\Bchtodstzdst$ decays
are pseudoscalar $\to$ vector vector (VV) transitions
described by three independent helicity amplitudes $A_0$, $A_{\parallel}$, and $A_{\perp}$~\cite{Dunietz}.  
The lack of knowledge of the true helicity amplitudes in
the $B \to VV$ final states contributes a systematic uncertainty to
the efficiency.  The dominant source of this effect originates
from the $p_T$-dependent inefficiency in reconstructing the low-momentum ``soft''
pions in the $\Dstarp$ and $\Dstarz$ decays, and the fact that the three helicity amplitudes contribute very
differently to the slow pion $p_T$ distributions.  
To estimate the size of this effect, MC samples are 
produced with a phase-space angular distribution model for the decay products.
Each event is then weighted by the
angular distribution for given input values of the helicity amplitudes
and phase differences.   The efficiency is then determined for a
large number of amplitude sets and the observed distributions in efficiencies
are used to estimate a systematic uncertainty.
For a given iteration, a random number, based on a uniform PDF,
is generated for each of the three parameters: $R_\perp, \alpha$, and $\eta$, where
\begin{equation}
\! R_\perp =  \frac{|A_{\perp}|^2}{|A_0|^2 + |A_{\parallel}|^2 + |A_{\perp}|^2}, \quad \alpha = \frac{|A_{0}|^2  - |A_{\parallel}|^2 }{|A_0|^2 + |A_{\parallel}|^2},
\end{equation}
and $\eta$ is the strong phase difference between $A_0$ and $A_{||}$. 
Since $R_\perp$ for \Bztodstdst has already been measured~\cite{babarDstDstCP}, 
a Gaussian PDF with mean and width fixed to the
measured values is used instead for that mode.
The events of the MC sample are weighted by the
corresponding angular distribution and the efficiency is determined
(after applying all selection cuts)
by fitting the \mes distribution and dividing by the number of generated events.
The procedure is repeated 1000 times for each $B \to VV$ sample.
The relative spread in efficiencies (rms divided by the mean)
is used to estimate the systematic
uncertainty due to a lack of knowledge of the true amplitudes.
\paragraph{Peaking background and crossfeed} The uncertainties on the peaking background vector $P_i$ and on the
efficiency matrix $\epsilon_{ij}$ are dominated by the available MC statistics.  The resulting uncertainties on each element
of the vector and matrix are propagated through to the branching fraction results via the formalism of Eq.~\ref{eq:bfs}.
\paragraph{Number of \BB} The number of \BB events in the full data sample, and the uncertainty on this number, are determined via
a dedicated analysis of charged track multiplicity and event shape~\cite{FoxW}.  The uncertainty introduces a systematic uncertainty of
1.1\% on each of the branching fractions.

\section{Measurement of {\boldmath \CP}-violating charge asymmetries}\label{sec:acps}

\begin{figure*}[p]
\begin{center}
\scalebox{0.53}{\includegraphics{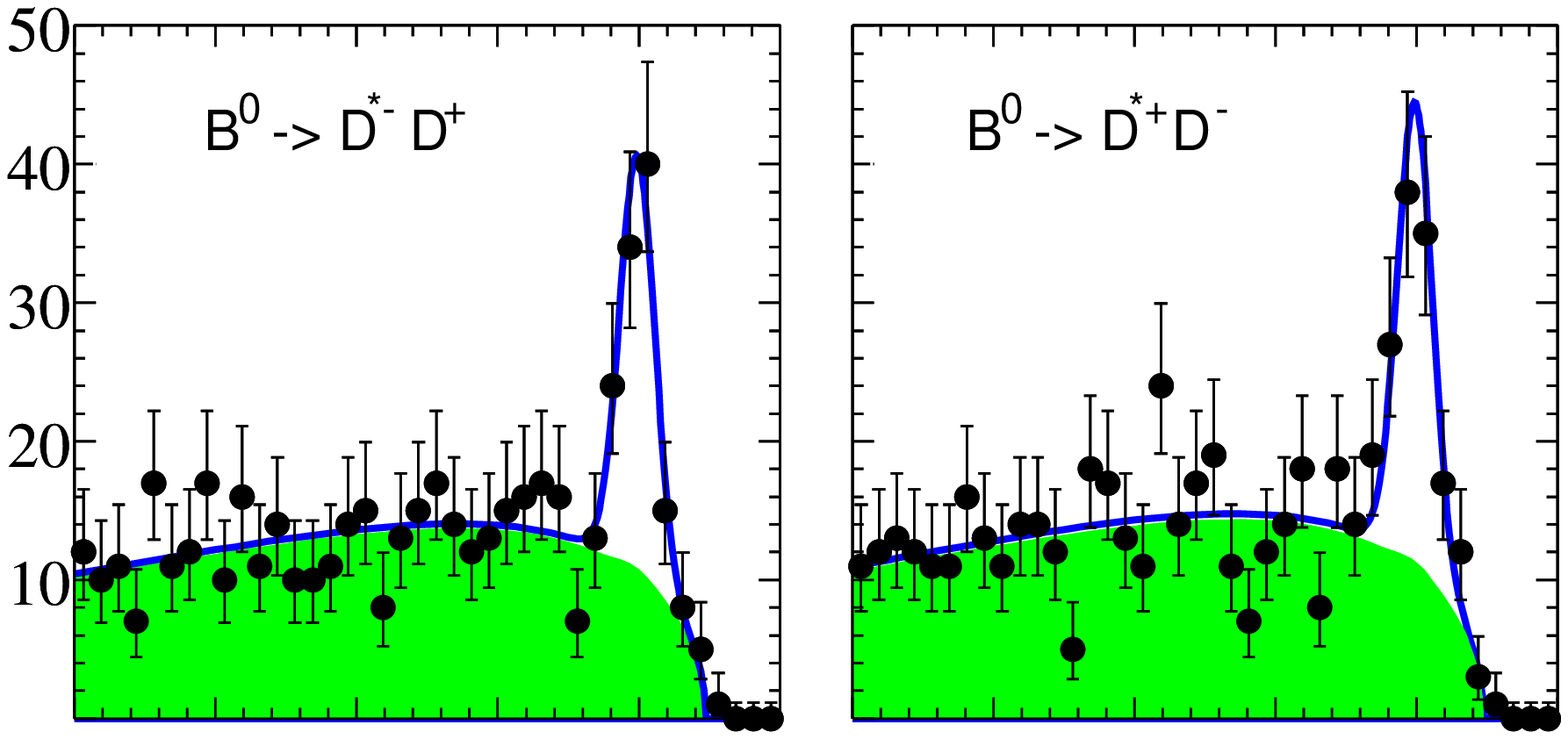}}

\vspace*{-1.2cm}
      
\scalebox{0.53}{\includegraphics{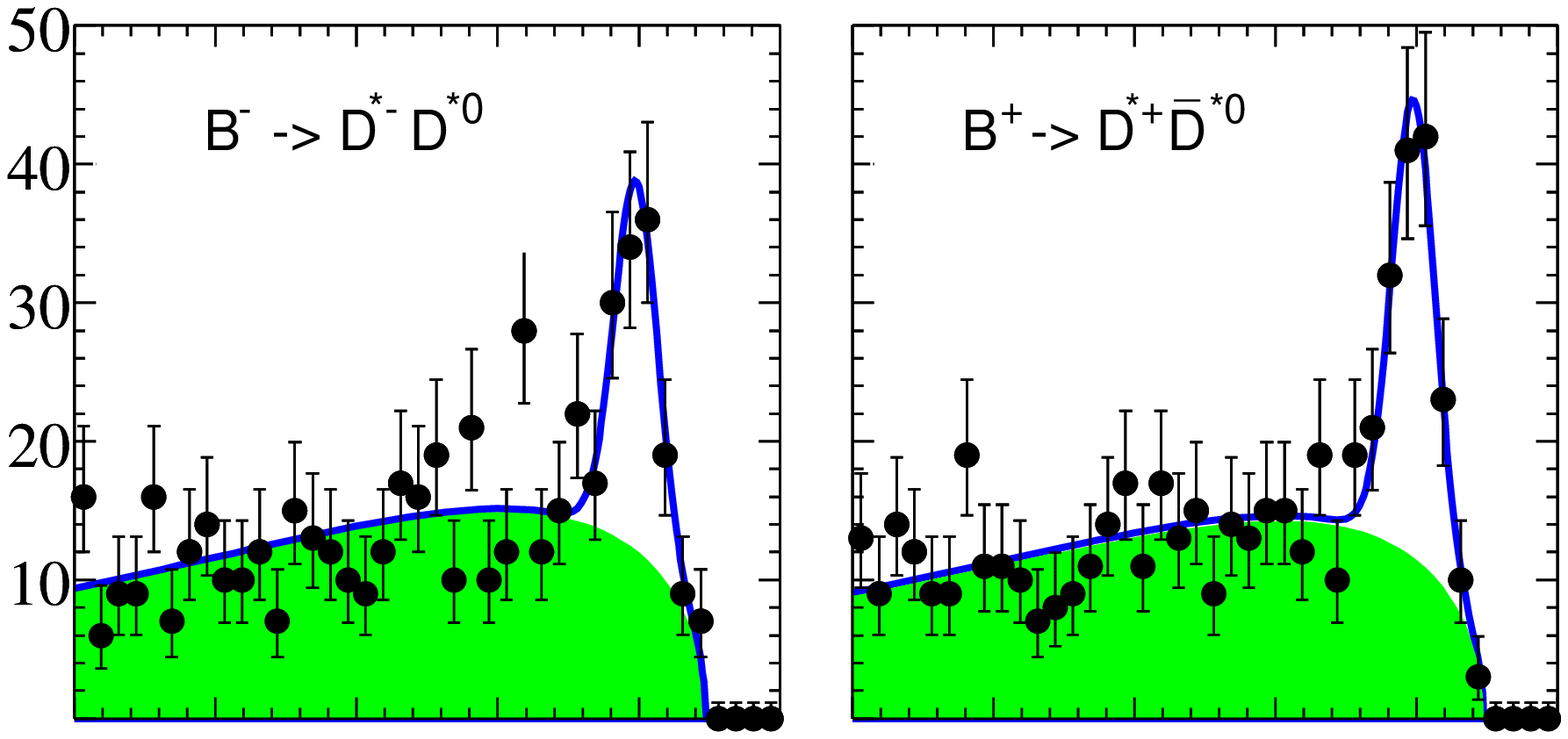}}

\vspace*{-1.2cm}

\scalebox{0.53}{\includegraphics{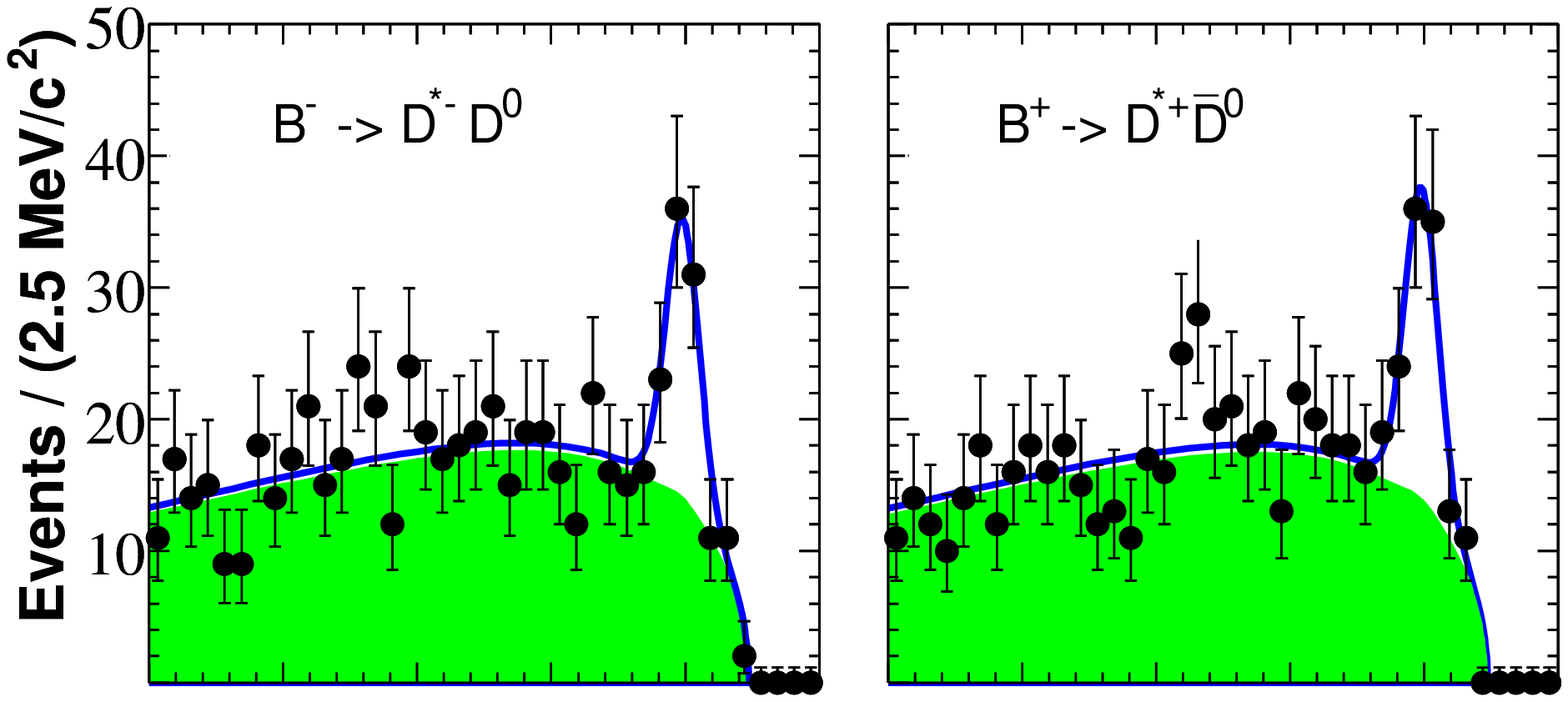}}

\vspace*{-1.2cm}
      
\scalebox{0.53}{\includegraphics{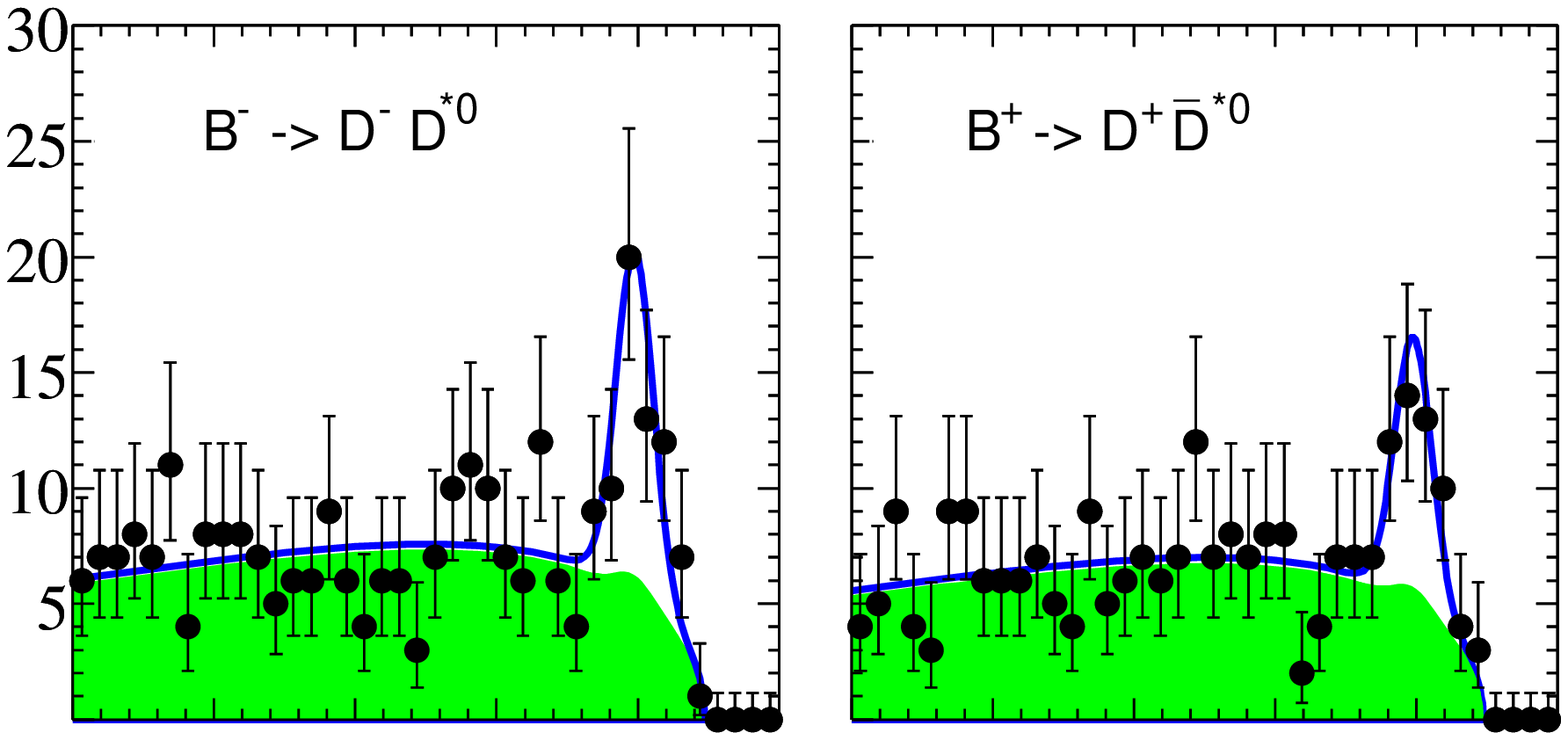}}

\vspace*{-1.2cm}

\scalebox{0.53}{\includegraphics{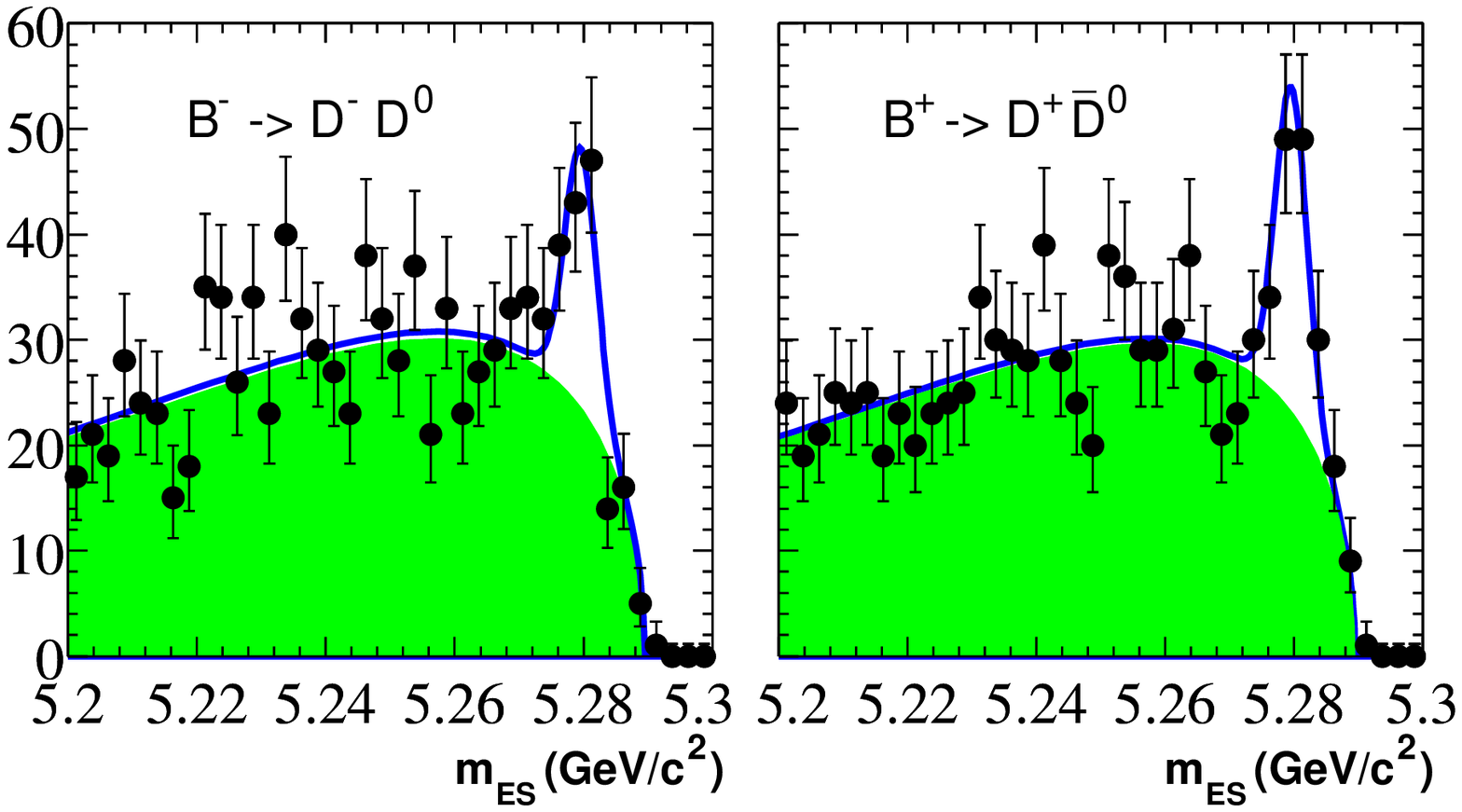}}

\vspace*{-0.5cm}

\end{center}
\caption{
\label{fig:data_acps}
Fitted distributions of \mes for the two conjugate states of each of the five relevant modes.
The error bars represent the statistical errors only.
The solid lines represent the fits to the data, and the shaded areas
the fitted background.
The raw asymmetries $\mathcal{A}$ are the normalized differences in the amount of signal between the members of each conjugate pair.
}
\end{figure*}

To obtain the charge asymmetries $\mathcal{A}_{\CP}$ (defined in Eq.~\ref{eq:acp}), we perform unbinned extended maximum likelihood fits to
the \mes distributions of the selected events in each of the four charged-$B$ decay modes $D^{*+}\Dstarzb$,
$D^{*+}\Dzb$, $D^{+}\Dstarzb$, $D^{+}\Dzb$, and their respective charge conjugates, and in the neutral-$B$ decay mode $\Dstarpm\Dmp$, using
Eq.~\ref{eq:argus} as the PDF for the combinatorial background for both charges in each pair.  The free parameters of each of the five fits individually
are: 1) the combinatorial background shape parameter
$\kappa$, 2) the total number of signal events, 3) the total number of background events,
and 4) the ``raw'' charge asymmetry $\mathcal{A}$.  Parameters 1 and 3 are considered (and thus constrained to be)
the same for both charge states in each mode; this assumption is validated in MC simulation of the background
as well as in control samples of $\Bz \to \Dstarm\rho^+$ and $\Bz \to \Dstarm a_1^+$ decays in data.  
The results of the fits are shown in Fig.~\ref{fig:data_acps}.
Two potentially biasing effects must be considered: there can be a asymmetry in the efficiencies for reconstructing
positively- and negatively-charged tracks, and peaking background and crossfeed between the modes can cause a small difference
between the measured (``raw'') asymmetry and the true asymmetry.  The former of those two effects is discussed in Sec.~VIII below. 
Regarding the latter, to obtain the charge asymmetries $\mathcal{A}_{\CP}$ from the ``raw''
asymmetries $\mathcal{A}$, very small corrections for peaking background and crossfeed between modes must be made.
Using the terminology of Eq.~\ref{eq:bfs}, and 
considering the branching fractions $\mathcal{B}_i$ to be sums of a ``$+$'' mode (with a \Bz or \Bp, containing a $\bar{b}$ quark, as the initial state) 
and a ``$-$'' mode (with a \Bzb or \Bm, which contain a $b$ quark, as the initial state): $\mathcal{B}_i \equiv \mathcal{B}_i^+ + \mathcal{B}_i^-$,
we have the two equations
\begin{equation}
\sum_j \epsilon_{ij} \mathcal{B}_j^{\pm} N_B = N^{\rm sig \pm}_i - P_i^{\pm}
\end{equation}
for the ``$+$'' and ``$-$'' states respectively, which imply
\setlength{\arraycolsep}{0.0mm}
\begin{eqnarray}
& & \mathcal{B}_i^- \pm \mathcal{B}_i^+ = \nonumber\\
& & \qquad \sum_j \epsilon^{-1}_{ij} [(N^{\rm sig -}_j - P_j^-) \pm (N^{\rm sig +}_j - P_j^+)] / N_B. \qquad 
\end{eqnarray}
\setlength{\arraycolsep}{1.0mm}
As 
\begin{equation}
\mathcal{A}_{\CP,i} \equiv \frac{\Gamma_i^{-} - \Gamma_i^{+}}{\Gamma_i^{-} + \Gamma_i^{+}} = \frac{\mathcal{B}_i^- - \mathcal{B}_i^+}{\mathcal{B}_i^- + \mathcal{B}_i^+}, 
\end{equation}
we have
\begin{equation}
\mathcal{A}_{\CP,i} = \frac{\sum_j \epsilon^{-1}_{ij} [(N^{\rm sig -}_j - P_j^-) - (N^{\rm sig +}_j - P_j^+)]}{\sum_j \epsilon^{-1}_{ij} [(N^{\rm sig -}_j - P_j^-) + (N^{\rm sig +}_j - P_j^+)]}.
\end{equation}
Since $N^{\rm sig}_j \equiv N^{\rm sig -}_j + N^{\rm sig +}_j$
and the ``raw'' asymmetry in a mode $\mathcal{A}_j \equiv \frac{N^{\rm sig -}_j - N^{\rm sig +}_j}{N^{\rm sig -}_j + N^{\rm sig +}_j}$, we have
\begin{equation}
\label{eqn:acpxfeed}
\mathcal{A}_{\CP,i} = \frac{\sum_j \epsilon^{-1}_{ij} [\mathcal{A}_j N^{\rm sig}_j - \mathcal{A}^P_j P_j]}{\sum_j \epsilon^{-1}_{ij} [N^{\rm sig}_j - P_j]}
\end{equation}
where $\mathcal{A}^P_j \equiv \frac{P_j^- - P_j^+}{P_j^- + P_j^+}$ are the charge asymmetries of the peaking backgrounds.
The total yields $N^{\rm sig}_j$, peaking backgrounds $P_j$, and efficiency matrix $\epsilon_{ij}$ are identical to those used for the branching fraction
measurements and are given in Tables~\ref{tab:bfsacps} and~\ref{tab:effs}.
The values $\mathcal{A}^P_j$ are nominally set to 0 and
are varied to obtain systematic uncertainties due to the uncertainty on the charge asymmetry of the peaking background (see Sec.~\ref{sec:acpsysts}).  
Thus, Eq.~\ref{eqn:acpxfeed} is used to determine the
final $\mathcal{A}_{\CP}$ values from the measured asymmetries, in order to account for the small effects due to 
peaking background and crossfeed between modes.
The measured $\mathcal{A}_{\CP}$ values 
are given in Table~\ref{tab:bfsacps}.  They are all consistent with zero, and their errors are dominated by statistical uncertainty.

\begin{table*}[t]
\caption{
Summary of the systematic uncertainties estimated for the $\mathcal{A}_{\CP}$ asymmetries, in \%.  
} 
\begin{center}
\begin{tabular}{lccccc}
\hline\hline
Systematics source                                 & $\Bz \to D^{*\pm}D^{\mp}$ & $\Bp \to \Dstarp\Dstarzb$ & $\Bp \to \Dstarp\Dzb$ & $\Bp \to \Dp\Dstarzb$ & $\Bp \to \Dp\Dzb$  \\
\hline
Slow pion charge asymmetry                         & $0.53$          &    $0.53$       &  $0.53$     &   ---       &   ---      \\
Charge asymmetry from other tracks                 & $0.80$          &    $0.80$       &  $0.80$     &  $0.80$     &  $0.80$    \\
Amount of peaking bkgd.                            & $0.06$          &    $0.42$       &  $0.19$     &  $0.64$     &  $0.53$    \\
$\mathcal{A}_{\CP}$ of peaking bkgd.               & $0.42$          &    $0.09$       &  $0.58$     &  $3.36$     &  $0.85$    \\
Crossfeed uncertainty                              & $0.00$          &    $0.00$       &  $0.01$     &  $0.31$     &  $0.00$    \\
\mes resolution uncertainty                        & $0.20$          &    $0.04$       &  $0.04$     &  $0.01$     &  $0.14$    \\
$B$ mass uncertainty                               & $0.20$          &    $0.37$       &  $1.38$     &  $1.38$     &  $0.53$    \\
Uncertainty in $\sqrt{s}$                          & $0.00$          &    $0.03$       &  $0.08$     &  $0.05$     &  $0.05$    \\
Potential fit bias                                 & $0.74$          &    $1.97$       &  $1.19$     &  $0.53$     &  $1.66$    \\
\hline 
TOTAL $\delta(\mathcal{A}_{\CP})$                  & $1.6$          &    $2.4$       &  $2.3$     &  $3.8$     &  $2.2$    \\
\hline \hline
\end{tabular}
\vspace{0.3cm} 
\label{tab:acpsysts}
\end{center}
\end{table*}

\setcounter{paragraph}{0}

\section{Systematic uncertainties on charge asymmetry measurements}\label{sec:acpsysts}

Table~\ref{tab:acpsysts} shows the results of our evaluation of the various sources of systematic uncertainty that are important for the $\mathcal{A}_{\CP}$ measurements.
\paragraph{Slow \pipm charge asymmetry} A charge asymmetry in the reconstruction efficiency of the low-transverse-momentum charged pions
from $\Dstarpm \to \Dz\pipm$ decays can cause a shift in $\mathcal{A}_{\CP}$ by biasing the rates of positively charged vs.~negatively charged decays for each
mode.  We estimate this systematic uncertainty by using data control samples of $\Bz \to \Dstarm X^+$ and  $\Bzb \to \Dstarp X^-$ decays, where
$X$ is either $\pi$, $\rho$, or $a_1$, and determining if there is an asymmetry in the number of \Dstarp vs.~\Dstarm reconstructed.  There
are two potential biases of this technique: 1) a charge asymmetry in tracks other than the slow charged pions, and 2) the presence of doubly-Cabibbo-suppressed
$\Bz(\Bzb) \to \Dstarmp X^{\pm}$ decays which could potentially introduce a direct-\CP-violating asymmetry between the two states in the control sample.  
Discussion of 1) is detailed in the paragraph below, and the rate of 2) has been determined in analyses such as Refs.~\cite{exclS2bg} and~\cite{partialS2bg} to
be of order $0.1\%$, well below the sensitivity for this measurement.  We combine the information from the control sample modes and determine
an uncertainty of 0.5\% for each $\mathcal{A}_{\CP}$ measurement for modes with a charged slow pion.
\paragraph{Charge asymmetry from tracks other than slow \pipm} Auxilliary track reconstruction studies place a stringent bound on 
detector charge asymmetry effects at transverse momenta above 200 \mevc.  Such tracking and PID systematic effects were studied in detail in
the analysis of $B \to \phi K^{*}$~\cite{phikstar}.  We assign a 0.2\% systematic per charged track, thus an overall systematic
of 0.4\% per mode (as the positively charged and negatively charged decays for each mode have, on balance, one positive vs.~one negative track respectively).
This systematic uncertainty is added linearly to the slow pion charge asymmetry systematic due to potential correlation.
\paragraph{Amount of peaking background} Peaking background can potentially bias $\mathcal{A}_{\CP}$ measurements in two ways:
1) a difference in the total amount of peaking background from the expected total amount can, to second order, alter the measured
asymmetry between the positively charged and negatively charged decays, 2) a more direct way for peaking background to alter the measured $\mathcal{A}_{\CP}$ would
be if the peaking background itself were to have an asymmetry between the amount that is reconstructed as positively charged and the amount reconstructed as negative.  
1) is discussed here; 2) is discussed in the paragraph below.  The systematic uncertainty due to the uncertainty on the total amount of 
peaking background in the five decays is determined via the formalism of Eq.~\ref{eqn:acpxfeed}.  Namely, the uncertainty is given by
\begin{equation}
\label{eqn:acppbsyst}
\delta\mathcal{A}_{\CP,i} = \frac{(\sum_j \epsilon^{-1}_{ij} \mathcal{A}_j N^{\rm sig}_j) \times \sqrt{\sum_j (\epsilon^{-1}_{ij})^2 (\delta P)^2_j}}
                                 {(\sum_j \epsilon^{-1}_{ij} [N^{\rm sig}_j - P_j])^2}
\end{equation}
where $(\delta P)_j$ are the uncertainties on the amount of peaking background (which are given, along with the other parameters in the equation, 
in Table~\ref{tab:bfsacps}).
\paragraph{$\mathcal{A}_{\CP}$ of peaking background} The systematic uncertainty due to the $\mathcal{A}_{\CP}$ of the peaking background is also determined using the formalism 
of Eq.~\ref{eqn:acpxfeed}.  Namely, the uncertainty is given by
\begin{equation}
\label{eqn:acppbacpsyst}
\!\!\!\!\delta^{\prime}\mathcal{A}_{\CP,i} = \frac{(\sum_j \epsilon^{-1}_{ij} \mathcal{A}_j N^{\rm sig}_j) \!\times\!\! \sqrt{\sum_j (\epsilon^{-1}_{ij})^2 (\delta A^P)^2_j P^2_j}}
{(\sum_j \epsilon^{-1}_{ij} [N^{\rm sig}_j - P_j]) (\sum_j \epsilon^{-1}_{ij} \mathcal{A}_j N^{\rm sig}_j)}.
\end{equation}
Investigation of the sources of the peaking background in these modes motivates a conservative choice of 0.68 for the $(\delta A^P)_j$ values.
\paragraph{Amount of crossfeed} The systematic error due to uncertainties in the amount of crossfeed between the modes is also 
determined via the formalism of
Eq.~\ref{eqn:acpxfeed}.  Namely, the uncertainty is given by
\begin{widetext}
\begin{eqnarray}
\label{eqn:acpxfeedsyst}
\delta^{\prime\prime}\mathcal{A}_{\CP,i} & = & \frac{\sqrt{\sum_{jk} \mathcal{A}_j N^{\rm sig}_j {\rm cov}(\epsilon^{-1}_{ij},\epsilon^{-1}_{ik}) \mathcal{A}_k N^{\rm sig}_k}}
                                                    {\sum_j \epsilon^{-1}_{ij} [N^{\rm sig}_j - P_j]}\\
 &   & \qquad - \quad \frac{(\sum_j \epsilon^{-1}_{ij} \mathcal{A}_j N^{\rm sig}_j) \times 
                                                      (\sqrt{\sum_{jk} [N^{\rm sig}_j - P_j] {\rm cov}(\epsilon^{-1}_{ij},\epsilon^{-1}_{ik})  [N^{\rm sig}_k - P_k]})}
                                                      {(\sum_j \epsilon^{-1}_{ij} [N^{\rm sig}_j - P_j])^2}.\nonumber
\end{eqnarray}
\end{widetext}
The covariance between the elements of the inverse efficiency matrix is obtained using the method of Ref.~\cite{matinverr}. 
The very small systematic uncertainty due to crossfeed is thus obtained using Eq.~\ref{eqn:acpxfeedsyst} and the amounts of crossfeed and their uncertainties that are given in 
Table~\ref{tab:effs}.
\paragraph{Uncertainty in \mes resolution, $B$ mass, and $\sqrt{s}$}
The uncertainties in \mes resolution and the beam energy $\sqrt{s}$ are
determined by varying these parameters within their fitted $\pm1 \sigma$ ranges and observing the resulting changes in $\mathcal{A}_{\CP}$.
The uncertainty in the reconstructed
$B$ mass can also have an impact on the fitted \mes distributions and thus on the fitted 
$\mathcal{A}_{\CP}$ values.  Varying the $B$ mass between the fitted value and the $\pm1 \sigma$ range of the nominal \Bz or \Bp invariant mass
allows the determination of the
resulting effect on the $\mathcal{A}_{\CP}$ values.
\paragraph{Potential fit bias} Uncertainties in the potential biases of the $\mathcal{A}_{\CP}$ fits are determined by performing the fits on large samples
of MC simulation of the signal decay modes and of \BB and continuum background decays.  All results are consistent with zero bias, and the 
uncertainties of the fitted asymmetries on the simulated data samples are conservatively assigned as systematic uncertainties from biases of the fits.

\section{Implications for {\boldmath $\gamma$}}\label{sec:gamma}

Information on the weak phase $\gamma$ may be obtained by
combining information from $B \to D^{(*)} \Dbar^{(*)}$ and $B \to D^{(*)}_s \Dbar^{(*)}$ branching
fractions, along with \CP asymmetry measurements in $B \to D^{(*)} \Dbar^{(*)}$,
and using an SU(3) relation between the $D^{(*)} \Dbar^{(*)}$ and $D^{(*)}_s \Dbar^{(*)}$ decays~\cite{DL,ADL}.
For this analysis, we assume that the breaking of SU(3) can be parametrized via the ratios of decay
constants $f_{D_s^{(*)}}/f_{D^{(*)}}$, which are quantities that can be
determined either with lattice QCD or from experimental measurements~\cite{lattice}.

In this model, one obtains the relation (for $\Bz \to \Dp\Dm$ and 
individual helicity states of $\Bz \to \Dstarp\Dstarm$):
\begin{equation}
\mathcal{A}_{ct}^2 = \frac{a_R \cos(2\beta + 2\gamma) - a_{\rm indir}\sin(2\beta + 2\gamma) - \mathcal{B}}{\cos 2\gamma - 1} 
\label{eqn:maineq}
\end{equation}
where
\begin{eqnarray}
\mathcal{B} & \equiv & \frac{1}{2}(|A^D|^2 + |\bar{A}^D|^2) = \\
  &        & \mathcal{A}_{ct}^2 + \mathcal{A}_{ut}^2 + 2\mathcal{A}_{ct}\mathcal{A}_{ut}\cos\delta\cos\gamma , \nonumber\\
a_{\rm dir} & \equiv & \frac{1}{2}(|A^D|^2 - |\bar{A}^D|^2) = \\
            &        & -2\mathcal{A}_{ct}\mathcal{A}_{ut}\sin\delta\sin\gamma ,\nonumber\\
a_{\rm indir} & \equiv & \Im[e^{-2i\beta}(A^{D})^{*}\bar{A}^D] = -\mathcal{A}_{ct}^{2}\sin 2\beta\nonumber\\
        & &     - 2\mathcal{A}_{ct}\mathcal{A}_{ut}\cos\delta\sin(2\beta + \gamma)\\
        & &     - \mathcal{A}_{ut}^2\sin(2\beta + 2\gamma) ,\nonumber
\end{eqnarray}
and
\begin{equation}
a_R^2 \equiv \mathcal{B}^2 - a_{\rm dir}^2 - a_{\rm indir}^2.
\end{equation}
$A^D$ and $\bar{A}^D$ represent amplitudes of a given $\Bz$ and $\Bzb \to D^{(*)+}D^{(*)-}$ decay respectively,
$\mathcal{B}$ represents the corresponding average branching fraction, and $a_{\rm dir}$ and $a_{\rm indir}$
represent the corresponding direct and indirect \CP asymmetries respectively.
The phases $\beta$ and $\gamma$ are the CKM phases and $\delta$ is a strong phase difference.
$\mathcal{A}_{ct} \equiv |(T + E + P_c - P_t - P_{EW}^C)V_{cb}^*V_{cd}|$ and
$\mathcal{A}_{ut} \equiv |(P_u - P_t -P_{EW}^C)V_{ub}^*V_{ud}|$ are the magnitudes of the combined $B \to D^{(*)} \Dbar^{(*)}$ decay amplitudes containing
$V_{cb}^*V_{cd}$ and $V_{ub}^*V_{ud}$ terms respectively, and the $T$, $P$, and $E$ terms are the
tree, penguin, and the sum of exchange and annihilation amplitudes respectively~\cite{DL}.  
One can directly measure the parameters $\mathcal{B}$, $a_{\rm dir}$, and $a_{\rm indir}$ using information from
$B \to D^{(*)} \Dbar^{(*)}$ decays; the parameter $\mathcal{A}_{ct}$ using information from $B \to D^{(*)}_s \Dbar^{(*)}$ decays;
and the weak phase $\beta$ can be obtained from the measurements of $\sin 2\beta$ based on $\Bz \to c\bar{c}\KS$
decays~\cite{HFAG} thus allowing for
solution of $\gamma$ (up to two discrete ambiguities) via Eq.~\ref{eqn:maineq}.
As the vector-pseudoscalar modes $\Bz \to \Dstarpm\Dmp$ are not \CP eigenstates,
a slightly more complicated analogue to Eq.~\ref{eqn:maineq} is needed for these modes~\cite{ADL}.  Measurement of $\mathcal{A}_{\CP}$ for $\Dstarpm\Dmp$
is also necessary to obtain information on $\gamma$ from the vector-pseudoscalar modes.

Using these relations, there are four variables besides $\beta$ for each $B \to D^{(*)} \Dbar^{(*)}$ decay for which to solve: $\mathcal{A}_{ct}$,
$\mathcal{A}_{ut}$, $\delta$, and $\gamma$.  The branching fraction and the direct and indirect
\CP asymmetries of the $B \to D^{(*)} \Dbar^{(*)}$ decay provide three measured quantities.  
The other measurement that can be used is the branching fraction of the corresponding
$B \to D^{(*)}_s \Dbar^{(*)}$ decay, by using the relation expressed in Eq.~\ref{eq:DsDrel}.

The values $a_{\rm indir}$ can, of course, only be measured in the neutral $B \to D^{(*)} \Dbar^{(*)}$ decays.  However, the charged
$B \to D^{(*)} \Dbar^{(*)}$ decays can supplement the neutral decays by adding information on $\mathcal{B}$ and $a_{\rm dir}$, assuming only
isospin symmetry between the charged and neutral modes.  Thus, information from the charged $B$ decay modes can assist the $\gamma$
determination.

SU(3)-breaking effects can distort the relation between $D^{(*)} \Dbar^{(*)}$ and $D^{(*)}_s \Dbar^{(*)}$ decays as expressed in 
Eq.~\ref{eqn:maineq}.
However, the SU(3)-breaking can be parametrized by the ratio of decay constants $f_{D_s^{(*)}}/f_{D^{(*)}}$, such that
the amplitude for $B \to D^{(*)}_s \Dbar^{(*)}$ decays
\begin{equation}
\mathcal{A}_{ct}^{\prime} = f_{D_s^{(*)}}/f_{D^{(*)}} \times \mathcal{A}_{ct}/\sin\theta_c
\label{eq:DsDrel}
\end{equation}
where $\theta_c$ is the Cabibbo angle~\cite{PDG2004} and the parentheses
around the asterisks correspond to the $B \to D^{(*)} \Dbar^{(*)}$ and $B \to D^{(*)}_s \Dbar^{(*)}$ decays that are used.
The theoretical uncertainty of this relation is determined to be 10\%~\cite{DL}.

We thus use the information from the vector-vector (VV) decays \Bz $\to D^{*+}D^{*-}$
and \Bp $\to D^{*+}\Dstarzb$ and pseudoscalar-pseudoscalar (PP) decays \Bz $\to D^{+}D^{-}$
and \Bp $\to D^{+}\Dzb$, as well as the vector-pseudoscalar (VP) decays
$\Bz \to D^{*\pm}D^{\mp}$, $\Bp \to \Dstarp\Dzb$, and $\Bp \to \Dp\Dstarzb$,
to form constraints on $\gamma$ using the method of Refs.~\cite{DL,ADL}.

To use the VV decays, we must make the assumption that the strong phases for
the $0$ and $\|$ helicity amplitudes are equal.  
The constraints from the PP decays require no such assumption.
The assumption of equal $0$ and $\|$ helicity amplitudes is theoretically supported by a QCD factorization argument described in~\cite{ADL}.  
Then, using Eq.~\ref{eqn:maineq},
we combine the \Bz $\to D^{*+}D^{*-}$ and \Bp $\to D^{*+}\Dstarzb$ branching
fractions and $\mathcal{A}_{\CP}$ information given above with measurements of the \Bz $\to D_s^{*-}D^{*+}$ and \Bp $\to D_s^{*+}\Dstarzb$ branching 
fractions~\cite{PDG2004},
measurements of the \Bz $\to D^{*+}D^{*-}$ time-dependent \CP asymmetries~\cite{babarDstDstCP,Belle2}, and the world-average values of $\sin
2\beta$~\cite{HFAG} and $\sin \theta_c$~\cite{PDG2004}.

We use a fast parametrized
MC method, described in Ref.~\cite{ADL}, to determine the confidence intervals for
$\gamma$.
We consider 500 values for $\gamma$, evenly spaced between 0 and
$2\pi$.  For each value of $\gamma$ considered, we generate 25000 
MC experiments, with inputs that are generated according to Gaussian distributions with widths
equal to the experimental
errors of each quantity.  
For each experiment, we generate random
values of each of the experimental inputs according to Gaussian
distributions, with means and sigmas according to the measured central
value and total errors on each experimental quantity.  We make the
assumption that the ratio $f_{D_s^*}/f_{D^*}$ is equal to $f_{D_s}/f_D
= 1.20 \pm 0.06 \pm 0.06$ \cite{lattice}, allowing for the additional 10\% theoretical uncertainty~\cite{DL}.
We then calculate the resulting values of $\mathcal{A}_{ct}$,
$a_{\rm dir}$, $a_{\rm indir}$, and $B$, given the generated random
values (based on the experimental values). When the quantities
$a_{\rm dir}$, $a_{\rm indir}$, and $B$, along with $\beta$ and the
value of $\gamma$ that is being considered, are input into
Eq.~(\ref{eqn:maineq}), we obtain a residual value for each
experiment, equal to the difference of the left- and right-hand sides
of the equation. 
Thus, using Eq.~\ref{eqn:maineq}, the 25000 trials per value of  $\gamma$ provide
an ensemble of residual values that are used to create a likelihood for $\gamma$ to be at that value, given the experimental inputs.
The likelihood, as a function of $\gamma$, can be
obtained from $\chi^2 (\gamma)$, where $\chi^2 \equiv (\mu/\sigma)^2$,
$\mu$ is the mean of the above ensemble of residual values, and
$\sigma$ is the usual square root of the variance.  The value of
$\chi^{2}(\gamma)$ is then considered to represent a likelihood which is
equal to that of a value $\chi$ standard devations of a Gaussian
distribution from the most likely value(s) of $\gamma$.
We define the ``exclusion level,'' as a function of the value of $\gamma$, as
follows: the value of $\gamma$ is excluded from a range at a given
C.L. if the exclusion level in that range of $\gamma$ values is 
greater than the given C.L.

We now turn to the VP decays. The method using VP decays shares the advantage
with PP decays that no assumptions on strong phases are required.
The disadvantage is that, as we will see, the constraints from the VP modes are weak.

We combine the information given above on the \Bz $\to D^{*\pm}D^{\mp}$, $\Bp \to \Dstarp\Dzb$, and $\Bp \to \Dp\Dstarzb$ branching fractions and $\mathcal{A}_{\CP}$
information
with measurements of the \Bz $\to D_s^{*-}D^{+}$, \Bz $\to D_s^{-}D^{*+}$, \Bp $\to D_s^{*+}\Dzb$, and \Bp $\to D_s^{+}\Dstarzb$ branching fractions~\cite{PDG2004},
measurements of the \Bz $\to D^{*\pm}D^{\mp}$ time-dependent \CP asymmetries~\cite{Babar2,belleDstDCP}, and the world-average values of $\sin
2\beta$~\cite{HFAG} and $\sin^2 \theta_c$~\cite{PDG2004}.
Similar to the MC $\gamma$ determination for the VV and PP modes,
we generate random values of each of the experimental inputs according
to Gaussian distributions, with means and sigmas according to the
measured central value and total errors on each experimental quantity.
We again obtain
a confidence level distribution as a function of $\gamma$.

\begin{figure*}[!t]
\begin{center}
\includegraphics[width=0.9\textwidth]{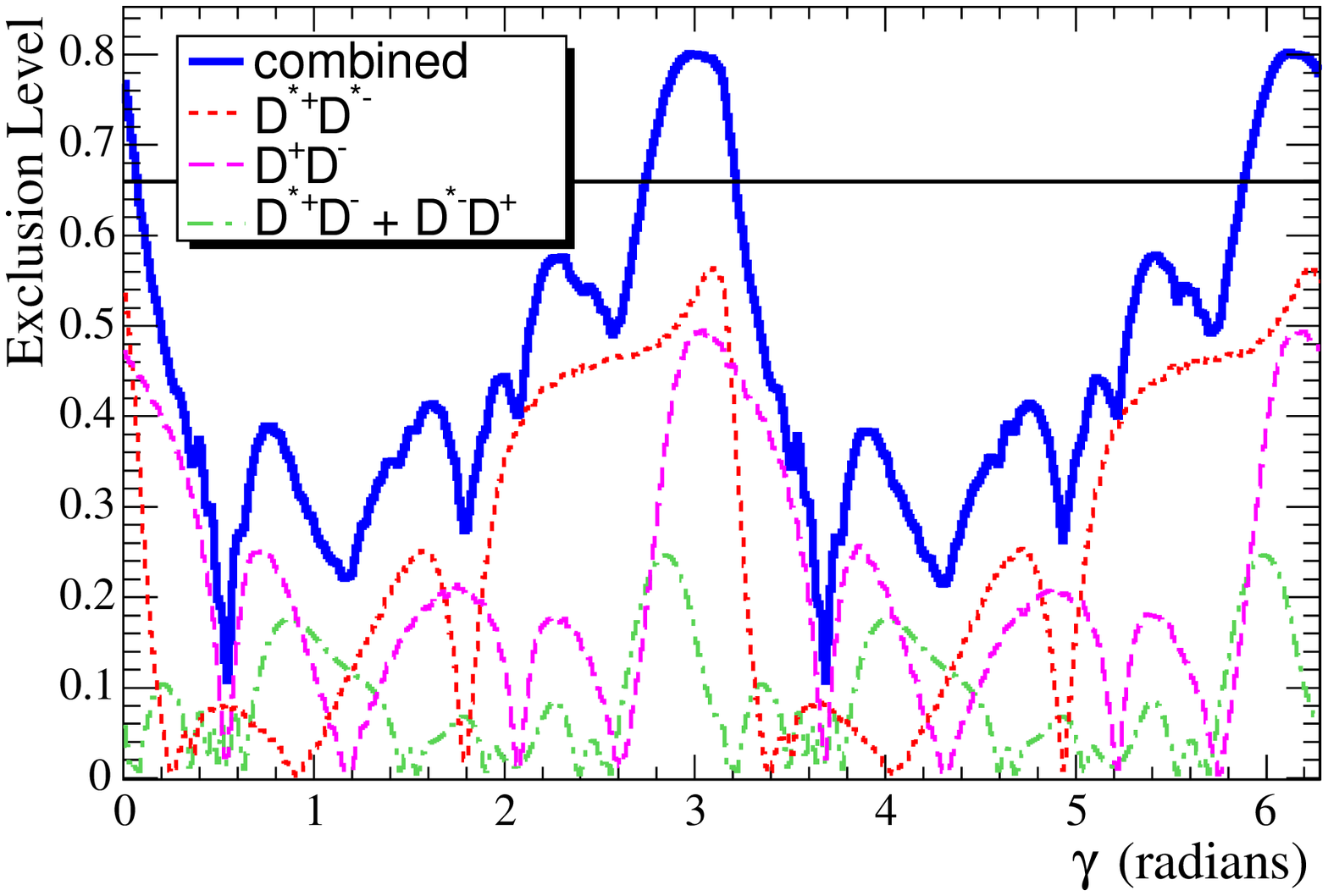}
\vspace*{-0.35cm}
\caption{ The measured exclusion level, as a function of $\gamma$,
from the combined information from vector-vector,
vector-pseudoscalar, and pseudoscalar-pseudoscalar modes.  The combined information implies that
$\gamma$ is favored to lie in the range $[0.07-2.77]$ radians (with a +0
or $+\pi$ radians ambiguity) radians at 68\% confidence level.}
\label{fig:chi2combined}
\end{center}
\end{figure*}

Finally, we can combine information from the VV, PP, and VP modes. The
resulting measured exclusion level as a
function of $\gamma$ from each of the three sets of modes, as well as from their combination,
is shown in Fig.~\ref{fig:chi2combined}. From the combined fit,
we see that $\gamma$ is favored to lie in the range
$[0.07-2.77]$ radians (with a +0 or $+\pi$ radians ambiguity) at 68\%
confidence level. This corresponds to $[4.1^{\circ}-158.6^{\circ}] (+0^{\circ} \mbox{ or } 180^{\circ})$.

These constraints are generally weaker than those found in Ref.~\cite{ADL} due to the fact
that the measured \CP asymmetry in \Bztodstdst has moved closer to the world-average \stwob,
with the newer \Bztodstdst measurements in Ref.~\cite{dstdst05prl}.  The closer this \CP
asymmetry is to \stwob, the weaker the resulting constraints are on $\gamma$, due to the fact
that the closeness of the \CP asymmetry to \stwob favors the dominance of the tree amplitude, rather than
the penguin amplitude whose phase provides the sensitivity to $\gamma$.
Although the constraints are not strong,
they contribute to the growing amount of information available on $\gamma$
from various sources.

\section{Conclusions}

In summary, we have measured branching fractions, upper limits, and charge 
asymmetries for all $B$ meson decays to $D^{(*)}\Dbar^{(*)}$.  The results are shown in Table~\ref{tab:bfsacps}.
This includes observation of the decay modes
\Bz $\to D^{+}D^{-}$ and \Bp $\to D^{*+}\Dstarzb$, evidence
for the decay modes \Bp $\to D^{+}\Dstarzb$ and \Bp $\to D^{+}\Dzb$
at $3.8 \sigma$ and $4.9 \sigma$ levels respectively,
constraints on \CP-violating charge asymmetries in the four decay modes \Bp $\to D^{(*)+}\Dbar^{(*)0}$,
measurements of (and upper limits for) the decay modes \Bz $\to D^{*0}\Dzb$
and \Bz $\to D^{0}\Dzb$, and improved branching fractions, upper limits, and charge asymmetries in all other $B \to D^{(*)}\Dbar^{(*)}$ 
modes. 
The results are consistent with theoretical expectation and (when available) previous measurements.
When we combine information from time-dependent \CP asymmetries in 
$\Bz \to D^{(*)+}D^{(*)-}$ decays~\cite{dstdst05prl,dstd05prl} and 
world-averaged branching fractions of $B$ decays to $D_s^{(*)}\Dbar^{(*)}$ modes~\cite{PDG2004} 
using the technique proposed in Ref.~\cite{DL} and implemented in Ref.~\cite{ADL}, we find the 
CKM phase $\gamma$ is favored to lie in the range $[0.07-2.77]$ radians (with a +0
or $+\pi$ radians ambiguity) at 68\% confidence level.

\section{Acknowledgments}\label{sec:Acknowledgments}

We are grateful for the 
extraordinary contributions of our \pep2\ colleagues in
achieving the excellent luminosity and machine conditions
that have made this work possible.
The success of this project also relies critically on the 
expertise and dedication of the computing organizations that 
support \babar.
The collaborating institutions wish to thank 
SLAC for its support and the kind hospitality extended to them. 
This work is supported by the
US Department of Energy
and National Science Foundation, the
Natural Sciences and Engineering Research Council (Canada),
Institute of High Energy Physics (China), the
Commissariat \`a l'Energie Atomique and
Institut National de Physique Nucl\'eaire et de Physique des Particules
(France), the
Bundesministerium f\"ur Bildung und Forschung and
Deutsche Forschungsgemeinschaft
(Germany), the
Istituto Nazionale di Fisica Nucleare (Italy),
the Foundation for Fundamental Research on Matter (The Netherlands),
the Research Council of Norway, the
Ministry of Science and Technology of the Russian Federation, and the
Particle Physics and Astronomy Research Council (United Kingdom). 
Individuals have received support from 
CONACyT (Mexico), the Marie-Curie Intra European Fellowship program (European Union),
the A. P. Sloan Foundation, 
the Research Corporation,
and the Alexander von Humboldt Foundation.


\begin{thebibliography}{99}
\bibitem{CC}
Except as noted otherwise, throughout the article we use a particle name to denote either member of 
a charge-conjugate pair.
\bibitem{Aleph1N}
The upper limit of $13 \times 10^{-3}$ is for the sum of the
$\Bp \to \Dstarp\Dzb$ and $\Bp \to \Dp\Dstarzb$
branching fractions.
\end{thebibliography}
\end{document}